\documentclass[aps,prb,twocolumn]{revtex4-1}

\usepackage{amsmath}
\usepackage{amssymb}
\usepackage{natbib}
\usepackage{graphicx}
\usepackage{bm}

\newcommand{\der}[2]{\frac{d{#1}}{d{#2}}}

\newcommand{\m}{\textbf{m}}

\newcommand{\al}[1]{\begin{align}#1\end{align}}
\newcommand{\matrixel}[3]{\left< #1 \vphantom{#2#3} \right| #2 \left| #3 \vphantom{#1#2} \right>}

\def\be{\begin{equation}}
\def\ee{\end{equation}}
\def\ba{\begin{eqnarray}}
\def\ea{\end{eqnarray}}
\def\rr{{\bf r}}

\def\a{\alpha}

\def\kk{{\bm k}}

\def\e{\epsilon}

\def\D{\Delta}
\def\p{\partial}

\def\pp{\bm{p}}
\def\rr{\bm{r}}
\def\qq{\bm{q}}
\def\vv{\bm v}

\def\EE{\bm{E}}

\def\d{\delta}

\def\bra{\langle}
\def\ket{\rangle}

\begin{document}

\title{Pancharatnam-Berry phase and kinetic magnetoelectric effect in a three-dimensional helical crystal (Te)}

\date{\today}
\author{C. \c{S}ahin, J. Rou, J. Ma, D. A. Pesin}
\affiliation{Department of Physics and Astronomy, University of Utah, Salt Lake City, UT 84112 USA}
\begin{abstract}
We study the kinetic magnetoelectric effect (current-induced magnetization including both the orbital and spin contributions) in three-dimensional conductors, specializing to the case of p-doped trigonal tellurium. We include both intrinsic and extrinsic contributions to the effect, which stem from the band structure of the crystal, and from disorder scattering, respectively. Specifically, we determine the dependence of the kinetic magnetoelectric response on the hole doping in tellurium, and show that the intrinsic and extrinsic effects dominate for low and high levels of doping, respectively. The results of this work imply that three-dimensional helical metals are promising for spintronics applications, in particular, they can provide robust control over current-induced magnetic torques.
\end{abstract}
\maketitle

\section{Introduction}

Control of magnetic degrees of freedom with electric current is one of the central subjects of spin electronics. In particular, the phenomenon of the current-induced magnetization, or the so-called kinetic magnetoelectric effect (KME), has received a lot of attention over the past several decades. For instance, in systems that lack inversion symmetry, magnetic torques due to the KME have been shown to efficiently control magnetization in various setups and materials\cite{Gambardella2011}. In this paper we describe the KME in three-dimensional noncentrosymmetric disordered conductors.

Experimentally, the typical setup for the observation of the KME has been based on a two-dimensional electron gas with strong spin-orbit coupling, whereby \emph{spin} magnetization would be induced by transport current. Such spin contribution to the current-induced magnetization was observed in Refs.~\onlinecite{Kato2004magnetization,silov2004current,Kato2005magnetization,Sih2005,Stern2006,Yang2006}. Three-dimensional materials offer new possibilities for the KME observation. Unlike the two-dimensional case, in three dimensions the orbital motion around the current flow is not quenched, hence there is a possibility of an orbital contribution to the KME\cite{Levitov,Yoda2015,Zhong2016}, which often is considerably stronger than the spin one. The experimental work on three-dimensional systems has been scarce. Notably, the intrinsic contribution to the KME was considered experimentally in p-doped tellurium in Ref.~\onlinecite{Farbshtein2012}, and a related phenomenon of current-induced optical activity in tellurium was observed in Ref.~\onlinecite{Ivchenko1979}.

The bulk of theoretical work on the KME related to the spin magnetic moment was done several decades ago in Refs.~\onlinecite{Ivchenko1978,VaskoPrima1979,Edelstein1990,Aronov1991} (see Ref.~\onlinecite{ganichev2016} for a review), while the orbital contribution to the KME was first considered in Ref.~\onlinecite{Levitov}.  That work also coined the term ``kinetic magnetoelectric effect'', which we find quite appropriate. Indeed, the predominance of the spin contribution to the KME in the literature is such that the term ``current-induced magnetization'' is firmly associated with the spin contribution to the effect. In this paper we therefore use the term ``kinetic magnetoelectric effect'' to alert the reader to the fact that in three-dimensional materials, there are also considerable orbital contributions to the KME.

Recently, with the advent of the field of Weyl materials, the interest in the KME has been revived, and it has been realized that in three dimensional materials the orbital effects are usually stronger than the spin ones\cite{Yoda2015,Zhong2016,Rou2017}. In particular, in Ref.~\onlinecite{Rou2017} it was shown that the theory of the KME in three-dimensional metals closely parallels the theory of the anomalous Hall effect~\cite{NagaosaReview}, in that the orbital contribution to the KME comes from both the intrinsic mechanisms, which are related to the band structure of a perfect crystal, as well as the extrinsic mechanisms, originating from the side jump and skew scattering from impurity .

In this work we consider the KME in a typical gyrotropic material - trigonal tellurium single crystals - focusing on both intrinsic and extrinsic contributions to the effect. For p-doped tellurium, we show that it is the intrinsic mechanism that determines the KME response at densities of holes up to a few of $10^{17}\textrm{cm}^{-3}$; for larger densities, the dominant mechanism is the skew scattering off impuritiets.

The rest of the paper is organized as follows. In Section~\ref{sec:results} we summarize the present state of the theoretical description of the KME. Section~\ref{sec:tellurium} is devoted to the description of an effective model for the tellurium band structure near the chemical potential. The results of Section~\ref{sec:tellurium} are used in Section~\ref{sec:KMEinTE} to calculate effective orbital magnetic moments, and finally the current-induced magnetization in tellurium. Some of the derivations pertaining to Section~\ref{sec:KMEinTE} are performed in Appendices~\ref{sec:appendixA}, \ref{sec:appendixB}, and \ref{sec:appendixC}. Finally, Section~\ref{sec:conclusions} contains a brief discussion of our results and conclusions.

\section{Kinetic magnetoelectric effect and Pancharatnam-Berry phase}\label{sec:results}

In this Section we summarize the current state of the KME theory in semimetals and doped semiconductors.

\subsection{Kinetic magnetoelectric effect}

We describe the magnetization response in the typical experimental situation of a transport current flowing through a sample. In the absence of the transport current, the sample is assumed to be time-reversal invariant. Under these conditions, the magnetization induced by the current can be written as\cite{footnoteHE}
\begin{align}\label{eq:KME}
  \bm M=-\frac{\tau}{\hbar}\int(d\kk)
 (\m^{\textrm{int}}_\kk+\m^{\textrm{sk}}_\kk+\m^{\textrm{sj}}_\kk) e\EE\cdot\p_\kk f^0_\kk,
\end{align}
where $\tau$ is the momentum relaxation time describing the impurity scattering in the relaxation time approximation, and $-\tau e \EE\cdot\p_\kk f^0_\kk/\hbar$ is the correction to the electrons' distribution function $f_\kk$ due to acceleration by the electric field $\EE$. $\m^{\textrm{int}}_\kk$ is the intrinsic contribution to the magnetic moment of electrons in a particular band, which contains both orbital and spin contributions, while $\m^{\textrm{sk}}_\kk$ and $\m^{\textrm{sj}}_\kk$ are the two parts of the orbital  magnetic moment induced by impurity scattering\cite{Rou2017}, which stem from the skew scattering and side-jump mechanisms, respectively.

For compactness of mathematical expressions, we omitted the band index on all quantities. In the case where several bands cross the Fermi level, the integral over quasi-wavevectors $\kk$ should be understood as
\begin{align}
  \int(d\kk)\to\sum_{n}\int(d\kk),\nonumber
\end{align}
where $n$ is the band index.

 The expression for $\m^{\textrm{int}}_\kk$ is given by
\begin{equation}\label{eq:magneticmoment}
\m^{\textrm{int}}_{\kk}=\frac {ie}{2\hbar} \bra \p_\kk u_{\kk}|\times (h_\kk-\e_{\kk})|\p_\kk u_{\kk}\ket+\frac{ge}{2m_0}\bra u_{\kk}|\bm s|u_{\kk}\ket,
\end{equation}
where $\bm s=\hbar \bm \sigma/2$ is the operator of electron's spin expressed through the vector of Pauli matrices, and $\bra u_{\kk}|\bm s|u_{\kk}\ket$ is its expectation value in a particular band. The orbital part of such intrinsic magnetic moment can be interpreted as self-rotation of a wave packet built from the wave functions of a given single band\cite{XiaoNiu}.

The kinetic contributions to the orbital magnetic moment, $\m^{\textrm{sk}}_\kk$ and $\m^{\textrm{sj}}_\kk$, come from the displacement of a wave packet center upon impurity scattering, which, in turn, is due to the skew scattering and side jump mechanisms. The theory of this contribution was developed in Ref.~\onlinecite{Rou2017}. Here we present a collection of the main results to facilitate further discussion in the context of tellurium.

Upon impurity scattering, an electron changes its quasimomentum from $\kk'$ to $\kk$, the scattering rate being $W_{\kk\kk'}=w_{\kk\kk'}\d(\e_{\kk}-\e_{\kk'})$.
The skew scattering contribution to the quasiparticle magnetic moment is related to the existence of an antisymmetric part of the scattering rate, $W_{\kk\kk'}\neq W_{\kk'\kk}$. In other words, the skew scattering is determined by $W^{\textrm{A}}_{\kk\kk'}=(W_{\kk\kk'}-W_{\kk'\kk})/2$. To the leading order in a weak disorder potential $V^{imp}$, $W^{\rm A}$ is given by\cite{Luttinger1958,Ghazali1972}
\begin{align}\label{eq:antisymmetricW}
  W^{\textrm{A}}_{\kk\kk'}=&-\frac{(2\pi)^2}{\hbar}n_i V^3 \int (d\kk''){\rm Im}(V^{imp}_{\kk\kk'}V^{imp}_{\kk'\kk''}V^{imp}_{\kk''\kk})\nonumber\\
  &\times \d(\e_\kk-\e_{\kk''})\d(\e_{\kk'}-\e_{\kk''}).
\end{align}
In this expression  $n_{i}$ is the impurity density, $V$ is the volume of the crystal, and $V^{imp}_{\kk\kk'}$ is the matrix element of the single-impurity potential between the full Bloch functions:
\begin{equation}\label{eq:impmatrel}
V^{imp}_{\kk\kk'} = \frac{1}{N}\int d\rr u^*_\kk(\rr)e^{-i\kk\rr} V^{imp}(\rr)
u_{\kk'}(\rr)e^{i\kk'\rr}.
\end{equation}
$N$ is the number of unit cells in the crystal.

The corresponding contribution to the orbital magnetic moment in the $\omega\tau\ll1$ limit is given by\cite{Rou2017}
\begin{equation}\label{eq:skmomentresults}
  \m^{\textrm{sk}}_\kk=e\tau^2\bm \int (d\kk')W^{\textrm{A}}_{\kk\kk'}
\vv_{\kk}\times\vv_{\kk'},
\end{equation}
where $\vv_\kk=\frac{1}{\hbar}\p_\kk \e_\kk$ is the usual group velocity of electrons with band energy $\e_\kk$.

In turn, the side jump contribution comes from the displacement, $\d \rr_{\kk\kk'}$, of a wave packet in the crystal's unit cell upon impurity scattering from $\kk'$ to $\kk$. For a  weak impurity  potential, $V^{imp}$, the side jump displacement is given by\cite{Belinicher1982,Sinitsyn2006}
\begin{align}\label{eq:sidejump}
  \d\rr_{\kk\kk'}=&i \bra u_{\kk}|\p_\kk u_{\kk}\ket-i\bra u_{\kk'}|\p_{\kk'} u_{\kk'}\ket-(\p_\kk+\p_{\kk'})\textrm{arg}(V^{imp}_{\kk\kk'}).
\end{align}
The side jump contribution to the quasiparticle magnetic moment is given by\cite{Rou2017}
\begin{equation}\label{eq:sjmomentresults}
  \m^{\textrm{sj}}_\kk=e\tau\int (d\kk')W^{\textrm{S}}_{\kk\kk'} \d \rr_{\kk\kk'}\times\vv_\kk.
\end{equation}

Equations~\eqref{eq:KME}, \eqref{eq:magneticmoment}, \eqref{eq:antisymmetricW}, \eqref{eq:skmomentresults}, \eqref{eq:sidejump}, and \eqref{eq:sjmomentresults} form the basis for our discussion of the KME in tellurium crystals.

\subsection{Relation to the Pancharatnam-Berry phase}

Equations \eqref{eq:antisymmetricW} and \eqref{eq:sidejump} for the skew scattering rate and the side jump displacement simplify in the case of a weak centrosymmetric impurity potential that varies slowly on the scale of the lattice constant. The resulting expressions also illustrate the geometric nature of these quantities, and serve as the basis for our subsequent considerations. We therefore briefly review them here.

In principle, Eqs.~\eqref{eq:antisymmetricW} and \eqref{eq:impmatrel} allow to calculate the antisymmetric part of the scattering rate for any impurity potential in any band structure. Practically, however, one has to make further assumptions in order to perform such a calculation without resorting to first-principle methods. As stated above, we assume that the impurity potential  $V^{imp}(\rr)$ is centrosymmetric, and varies slowly on the scale of the lattice constant, and that the Fermi pockets in tellurium are small, such that for relevant $\kk$'s one has $ka\ll1$, where $a$ is the typical size of the unit cell. Under these assumptions, the skew scattering rate~\eqref{eq:antisymmetricW} and the side jump~\eqref{eq:sidejump} can be expressed via the Pancharatnam-Berry phase $\Phi_{\kk,\kk',\kk''}$ in the electronic band structure\cite{Sinitsyn2006}:
\begin{align}\label{eq:PBphase}
  \Phi_{\kk,\kk',\kk''}&={\rm arg}\left[\bra u_{\kk}|u_{\kk'}\ket \bra u_{\kk'}|u_{\kk''}\ket\bra u_{\kk''}|u_{\kk}\ket\right].
\end{align}

To show this, we note that the matrix element~\eqref{eq:impmatrel} of a smooth impurity potential can be written as
\begin{align}
  V^{imp}_{\kk\kk'} = \frac{1}{V}V^{imp}(\kk-\kk') \bra u_\kk|u_{\kk'}\ket.
\end{align}
Here $V^{imp}(\kk-\kk') $ is the usual Fourier transform of the impurity potential:
\begin{align}
  V^{imp}(\qq)=\int d\rr e^{-i\qq\rr} V^{imp}(\rr).
\end{align}
For a centrosymmetric potential, $V^{imp}(\rr)=V^{imp}(-\rr)$, one can easily show that $\textrm{Im}[ V^{imp}(\qq)]=0$, and the antisymmetric part of the scattering rate is reduced to
\begin{widetext}
\begin{align}\label{eq:antisymmetricWPBphase}
  &W^{\textrm{A}}_{\kk\kk'}\approx-\frac{(2\pi)^2}{\hbar}n_i V^{imp}(\kk-\kk')V^{imp}(\kk'-\kk'')V^{imp}(\kk''-\kk)  \int (d\kk'')\d(\e_\kk-\e_{\kk''})\d(\e_{\kk'}-\e_{\kk''}) {\rm Im}\left[\bra u_\kk|u_{\kk'}\ket \bra u_{\kk'}|u_{\kk''}\ket\bra u_{\kk''}|u_\kk\ket\right].
\end{align}
It is obvious that the quantity
\begin{align}
  \Upsilon_{\kk,\kk',\kk''}&={\rm Im}\left[\bra u_{\kk}|u_{\kk'}\ket \bra u_{\kk'}|u_{\kk''}\ket\bra u_{\kk''}|u_{\kk}\ket\right],
\end{align}
which determines the skew scattering rate for a centrosymmetric impurity is non-zero only if the Pancharatnam-Berry phase~\eqref{eq:PBphase} is nonzero.

The side jump displacement~\eqref{eq:sidejump} can also be expressed through the Pancharatnam-Berry phase for weak centrosymmetric impurities. In this case
\begin{align}\label{eq:sidejumpPBP}
  \d\rr_{\kk\kk'}=&i \bra u_{\kk}|\p_\kk u_{\kk}\ket-i\bra u_{\kk'}|\p_{\kk'} u_{\kk'}\ket-(\p_\kk+\p_{\kk'})\textrm{arg}\bra u_\kk|u_{\kk'}\ket.
\end{align}
By noting that $\textrm{arg}(z)=\frac{1}{2i}\log z/z^*$ for a complex number $z$, and that $(\bra u_\kk|u_{\kk'}\ket)^*=\bra u_{\kk'}|u_{\kk}\ket$, one can obtain
\begin{equation}\label{eq:sidejumpPBphase}
  \d\rr_{\kk\kk'}=\lim_{\kk''\to\kk}\p_{\kk''}\Phi_{\kk,\kk',\kk''}+\lim_{\kk''\to\kk'}\p_{\kk''}\Phi_{\kk,\kk',\kk''}.
\end{equation}

Expressions~\eqref{eq:magneticmoment}, \eqref{eq:antisymmetricWPBphase}, and \eqref{eq:sidejumpPBphase} establish the geometric nature of the KME.
\end{widetext}
\section{Low-energy model for tellurium}\label{sec:tellurium}

\begin{figure}
  \centering
  \includegraphics[width=3in]{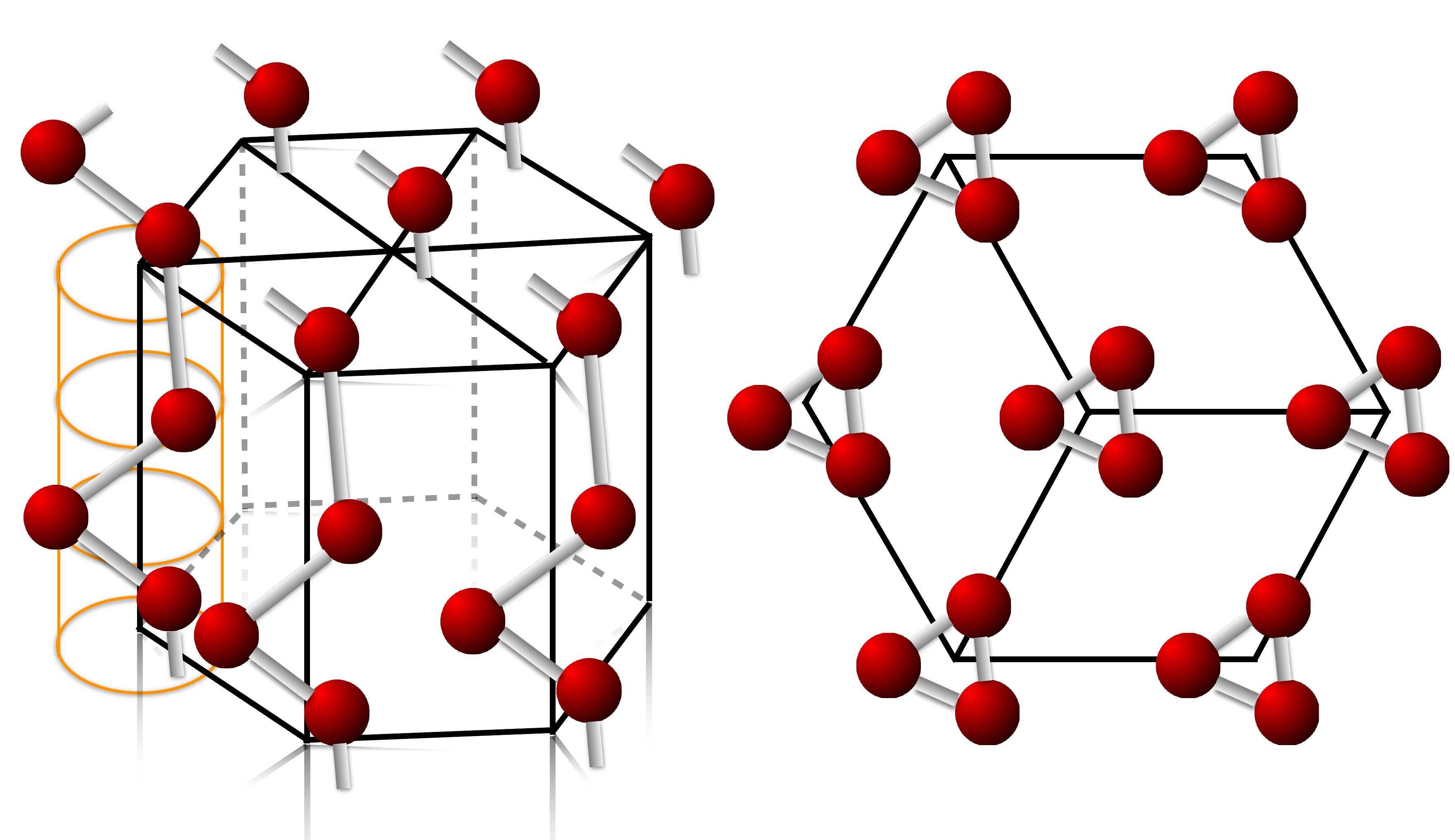}
  \caption{(Color online) tellurium crystal structure. Atomic helices wind around the vertical edges of the hexagonal prism representing the unit cell of the crystal. Top view of the structure is depicted on the right side.}\label{fig:Te}
\end{figure}
\subsection{The Effective Hamiltonian}\label{sec:effham}
In this section we explain the derivation of an effective Hamiltonian for trigonal tellurium which is an enantiomorphic crystal with a chiral point group $D_3$. Tellurium has motivated early works on the natural optical activity of semiconductors~\cite{Nomura1960,IvchenkoTe}. Recently, the interest in Te has been revived in various studies on the subject of topological insulators and metals, since it has been proposed to become a topological insulator or Weyl metal under pressure.~\cite{Agapito2013,Hirayama2015,Nakayama2017}

\begin{figure}
  \centering
  \includegraphics[width=3in]{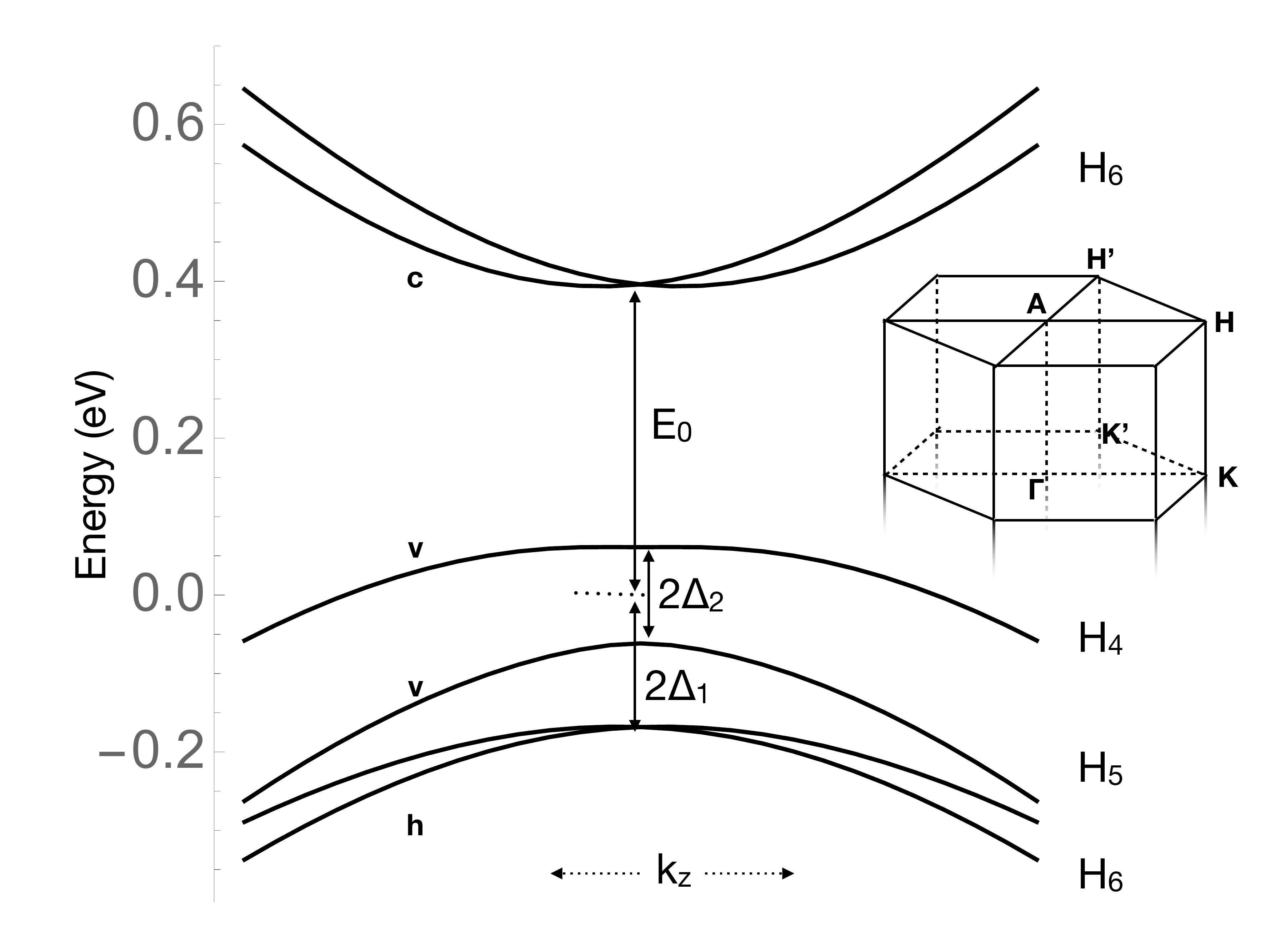}
  \caption{Band structure of tellurium along the H-K line in the Brillouin zone. The six bands closest to the chemical potential are shown (two conduction and four valence bands), together with their group representations.  The irreducible representations of the double group at the H-point consist of 2-dimensional H$_6$ representation for conduction and hole bands, as well as one-dimensional H$_4$ and H$_5$ representations for the upper and lower valence bands respectively. \cite{Doi1970} Inset: the Brillouin zone of tellurium, with relevant symmetry points indicated. }\label{fig:dispersions}
\end{figure}

The Brillouin zone of tellurium is a hexagonal prism, and the fundamental gap appears at H and H$'$ points at corners of the prism, located at $(\pm\frac{4\pi}{3a}, 0, \frac{\pi}{c})$ and their equivalents, see Fig.~\ref{fig:dispersions}. These points are related to each other by time-reversal symmetry, much like the K and K$'$ points in graphene. In the effective Hamiltonian we include the two lowest degenerate conduction bands and the four highest valence bands with a band gap of $E_g=0.335 eV$. The topmost valence band is separated from the second one as a result of the spin-orbit interaction by an energy of $2\Delta_2$ and the second valence band is separated from the lower two degenerate bands by $2\Delta_1$, where $\Delta_1=0.053$ eV and $\Delta_2=0.060$ eV, see Fig.~\ref{fig:dispersions}. Thus, the effective Hamiltonian acts in the basis in the reduced space of these six bands: $\psi_{c1}$, $\psi_{c2}$, $\psi_{v1}$, $\psi_{v2}$, $\psi_{h1}$, and $\psi_{h2}$ ordered in descending energy. We obtain the effective Hamiltonian perturbatively:
\begin{equation}
H_{eff}=H^{(0)} + H^{(1)} = H(\kk_0)+\der{H}{\kk}(\kk -\kk_0).
\end{equation}
We define $\bm \kappa=\kk - \kk_0$ for simplicity and in this picture the final form of the matrix elements of the effective Hamiltonian can be written as
\al{\label{eq:effham}
H_{nm}&=E_n\delta_{nm}+C^{nm}_i\kappa_i \nonumber
\\&-\frac{1}{2} \sum_k \frac{C^{nk}_{i}\kappa_i C^{km}_{j}\kappa_j}{E_k(\kk_0)-E_n(\kk_0)}
-\frac{1}{2} \sum_k \frac{C^{nk}_{j}\kappa_j C^{km}_{i}\kappa_i}{E_k(\kk_0)-E_m(\kk_0)},
}
where summation is over the states that are outside the reduced basis and we define
\begin{equation}\label{eq:cfactor}
C^{nm}_i=\matrixel{u_n(\kk_0)}{\partial_{\kk_i}H}{u_m(\kk_0)}.
\end{equation}
It is sufficient to know the wave functions at $\kk_0$ along with the total Hamiltonian to reconstruct the effective Hamiltonian. There have been many attempts to theoretically determine the parameters describing the band structure of tellurium, such as a $\kk \cdot \pp$ approach using group theory \cite{Doi1970}, which only included top four valence bands; a pseudopotential method \cite{Joannopoulos1975}; a simple tight-binding model with only nearest neighbor interactions \cite{Reitz1957}; and more recent tight-binding models in non-orthogonal basis \cite{HartMann1972, Li2013}; as well as purely group-theoretical phenomenological treatments\cite{}. In order to compute low energy parameters in Eq.~\eqref{eq:cfactor} we use the tight-binding Hamiltonian parameterized in Ref. \onlinecite{Molina2003} which we construct using atomic \textit{s} and \textit{p} orbitals with three helically oriented tellurium atoms per unit cell as shown in Fig.~\ref{fig:Te}. We include up to second nearest neighbors and use an orthogonal sp$^3$ basis. The resulting 24-band tight-binding model is adequate to obtain the
characteristics of the electronic band structure and effective masses around the Fermi energy. The spin-orbit coupling of 0.45 eV which is not included in the original parametrization \cite{Molina2003} is added later and the band gap is fitted to the experimental value of 0.335 eV. The inclusion of the spin-orbit interaction correctly splits the top two valence bands and accurately reproduces the famous ``humps'' around the valence band edge along the k$_z$ direction, Fig.~\ref{fig:dispersions}.

The low-energy model of tellurium describes the two conduction and four valence bands closest to the Fermi energy. For clarity, we will refer to the lowest two valence bands as ``hole'' ones - they only participate as intermediate states in the perturbation theory developed in the subsequent Sections, hence we would like to clearly separate them from the upper valence band that determines the DC transport in the system at low temperatures. The $6\times6$ effective Hamiltonian describing the low-energy states can be split into nine of $2\times2$ blocks:
\begin{align}\label{eq:3x3hamiltonian}
h_\kk=\left(
\begin{array}{ccc}
h_{cc}& h_{cv} & h_{ch} \\
h_{vc}& h_{vv} & h_{vh} \\
h_{hc}& h_{hv} & h_{hh} \\
\end{array}
\right).
\end{align}
In this Hamiltonian, indices $c,v,h$ pertain to the conduction, valence, and hole bands, respectively. $h$-matrices with coincident indices  -- $h_{cc},h_{vv},h_{hh}$ --  describe the bands without interband interaction, and the matrices with distinct indices (\textit{e.g.} $h_{cv}$) correspond to band interaction.

Explicitly, the Hamiltonian for the six bands closest to the Fermi level is given by

\begin{widetext}
\begin{align}\label{eq:matrixham}
h_\kk=\left(
\begin{array}{cc|cc|cc}
E_0+\beta_c k_z+\lambda_c & \delta_{1c} k_+^2+Ak_- & 0 & P_1 k_+ &  P_3k_z & P_2 k_- \\
 \delta_{1c} k_-^2+ A^*k_+ &E_0-\beta_c k_z+\lambda_c & P_1 k_- & 0 &  P_2 k_+ & - P_3 k_z \\
 \hline
 0 &  P_1k_+ & \beta_v k_z+\lambda_v & \Delta_2 & 0 & P_4 k_+ \\
  P_1 k_-& 0 & \Delta_2 & - \beta_v k_z +\lambda_v &  P_4k_-  & 0 \\
 \hline
  P_3k_z &P_2k_- & 0 & P_4k_+ &  \beta_h k_z-2 \Delta_1+\lambda_h & \delta_{1h} k_+^2+ B k_- \\
  P_2k_+ & -P_3k_z &  P_4k_- & 0 & \delta_{1h} k_-^2+ B^*k_+ & -k_z \beta_h-2 \Delta_1+\lambda_h \\
\end{array}
\right).
\end{align}
\end{widetext}
The vertical and horizontal lines in the Hamiltonian matrix separate $2\times2$  blocks introduced in Eq.~\eqref{eq:3x3hamiltonian}. The form of this Hamiltonian depends on a particular choice of basis vectors for (degenerate) bands at the H-point. We followed the form chosen in Ref.~\onlinecite{IvchenkoTe}.

The parameters in the Hamiltonian~\eqref{eq:matrixham} are as follows: $A=(\alpha_c +i\delta_{2c}k_z)$, $B=(\alpha_h+i  k_z \delta_{2h})$, $k_{\pm}=k_x \pm i k_y$, and $\lambda_i =\frac{\hbar^2}{2m^{0,i}_\parallel}k_z^2 +\frac{\hbar^2}{2m^{0,i}_\perp }k_\perp^2$ while $k_\perp^2=k_x^2+k_y^2$ and index $i=c,v,h$ represents the conduction, valence, and hole bands. Note that $m^{0,i}_{\perp,\parallel}$ are mass parameters without contributions from the band interaction effects (hence superscript `0'). In the Table~\ref{tab:bandparameters} we give the observed values of masses, which directly determine the band dispersions near the H-point.

 It should be noted that the threefold rotation symmetry of tellurium allows additional diagonal matrix elements in band-interaction blocks $h_{cv}$, $h_{vc}$, $h_{hv}$, and $h_{vh}$. However, explicit calculation shows that these terms are at least an order of magnitude smaller than the ones given, and hence are not included.

The parameters that enter into the Hamiltonian~\eqref{eq:matrixham} can be easily calculated from the effective Hamiltonian, $H$, derived from the tight-binding model, such as:
\al{
\beta_{v,c,h}&=\matrixel{\psi_{v1,c1,h1}(\kk_0)}{\partial_{\kk_z}H}{\psi_{v1,c1,h1}(\kk_0)}\\
\alpha_{c,h}&=\matrixel{\psi_{c1,h1}(\kk_0)}{\partial_{\kk_x}H-\partial_{\kk_y}H}{\psi_{c2,h2}(\kk_0)}\\
P_1&=\matrixel{\psi_{c1}(\kk_0)}{\partial_{\kk_x}H+i\partial_{\kk_y}H}{\psi_{v2}(\kk_0)}\\
P_2&=\matrixel{\psi_{c1}(\kk_0)}{\partial_{\kk_x}H-i\partial_{\kk_y}H}{\psi_{h2}(\kk_0)}\\
P_3&=i\matrixel{\psi_{c1}(\kk_0)}{\partial_{\kk_z}H}{\psi_{h2}(\kk_0)}\\
P_4&=\matrixel{\psi_{v1}(\kk_0)}{\partial_{\kk_x}H-i\partial_{\kk_y}H}{\psi_{h2}(\kk_0)}\\
P_5&=i\matrixel{\psi_{v1}(\kk_0)}{\partial_{\kk_z}H}{\psi_{h2}(\kk_0)}
}.
\begin{table}[h]
\centering
\begin{tabular}{|cl|cl|}
\hline
$\beta_v$: & 2.4 eV \AA      & $P_5$:             & 0 \\
$\beta_c$: & -0.76 eV \AA   & $\alpha_c$:      &0.067 eV \AA\\
$\beta_h$: & 0.27 eV \AA    &$\alpha_h$:       &0.047 eV \AA\\
$P_1$:       &  3.351 eV \AA & $\delta_{1c}$:  & 0.368 eV \AA$^2$ \\
$P_2$:       &  3.543 eV \AA &$\delta_{2c}$:   & 2.268 eV \AA$^2$\\
$ P_3$:      &  2.003 eV \AA & $\delta_{1h}$:  & 0.138 eV \AA$^2$\\
$P_4$:       &  1.933 eV \AA & $\delta_{2h}$:  &0.880 eV \AA$^2$\\
$\frac{\hbar^2}{2m_\parallel ^v}$: & -36.4 eV \AA$^2$ & $\frac{\hbar^2}{2m_\perp ^v}$: & -32.6 eV \AA$^2$  \\
$\frac{\hbar^2}{2m_\parallel ^c}$: & 52.9 eV \AA$^2$ & $\frac{\hbar^2}{2m_\perp ^c}$: & 57.5 eV \AA$^2$  \\
$\frac{\hbar^2}{2m_\parallel ^h}$: & -34.3 eV \AA$^2$ & $\frac{\hbar^2}{2m_\perp ^h}$: & -46.0 eV \AA$^2$  \\
\hline
\end{tabular}
\caption{\label{tab:bandparameters}
Band parameters for the low-energy Hamiltonian~\ref{eq:matrixham}}
\end{table}

We obtain band parameters that are summarized in Table \ref{tab:bandparameters}. Terms that are quadratic in $\kk$  are calculated from the effective masses along the parallel (c-axis) and perpendicular (in-plane) directions. Values of $P$'s in the table are already multiplied by $\hbar /m_0$, $m_0$ being the bare mass of the electron. Some of these parameters, such as $\beta_v$,  $\beta_c$, $\hbar^2 /2m_\parallel ^v$, and $\hbar^2 /2m_\perp ^v$ have been experimentally extracted and our values are in an excellent agreement with these studies \cite{Ivchenko1979, Bresler1970, Farbshtein2012, Shinno1973}.

The resulting  band dispersions of the tellurium conduction, valence, and hole bands are shown in Fig.~\ref{fig:dispersions}.

\section{KME in tellurium}\label{sec:KMEinTE}

In this Section we calculate the orbital magnetic moments of quasiparticles in tellurium, as well as the corresponding contributions to the KME. For realistic hole doping levels and at low temperatures, the Fermi level is located in the upper valence band\cite{Bresler1970}. Thus we restrict our attention to the quantities pertaining to that band only. The key observation in this regard is that the Pancharatnam-Berry phase and the orbital magnetic moment for this band come from its interaction with the conduction and hole bands (see below). This is unlike the conduction and hole bands themselves, for which these geometric quantities are nonzero even without band interaction. Therefore, the intrinsic magnetic moments for these bands, as well as the related physical observables, can be calculated as for isolated Rashba-type two-band band structures, similar to the treatment in Ref.~\onlinecite{MaPesin2015}, but this task is not further pursued in this paper.

\subsection{Effective orbital moments}\label{sec:intrinsic}

The main goal of this paper is a calculation of the magnetization response to a transport current flowing along the optical axis of a tellurium crystal. Such magnetization is determined by $z$-components of effective orbital magnetic moments. Therefore, below we restrict ourselves to the calculation of these components only. The other two can be found analogously, if need be.

\subsubsection{Intrinsic magnetic moment in the upper valence band}
 We start with the intrinsic orbital magnetic moment of quasiparticles for band $n$:
\begin{align}\label{eq:intrinsicm}
\mathbf{m}^{\mathrm{orb}}_n(\kk)=\frac{ie}{2\hbar}\bra\p_\kk u_n(\kk)|\times(h_\kk-\e_\kk)|\p_\kk u_n(\kk)\ket,
\end{align}
where $|u_n\ket$ is the periodic part of the Bloch wave function of band $n$, and $e<0$ is the charge of the electron.

Inserting a resolution of identity into Eq.~\eqref{eq:intrinsicm}, and using a well-known relation
\begin{align}
\bra u_m|\p_\kk h_\kk|u_n\ket=(\e_n-\e_m)\bra u_m|\p_\kk u_n\ket,\quad m\neq n,\nonumber
\end{align}
we arrive at an expression for the intrinsic orbital magnetic moment that does not require differentiating band eigenstates, and is more convenient for calculations:
\begin{align}\label{eq:mformula}
m_{n,a}^{\mathrm{orb}}=\frac{ie}{2\hbar}\epsilon_{abc}\sum_{m\neq n}\frac{\bra u_n|\p_b h_\kk|u_m\ket\bra u_m|\p_c h_\kk|u_n\ket}{\e_n-\e_m}.
\end{align}
Here $\p_a$ is shorthand notation for $\frac{\p}{\p k_a}$.

To evaluate the magnetic moment \eqref{eq:mformula}, we make several simplifications. We neglect the spin splittings of the conduction and hole bands. This is justified by the small value of $\beta_h$, as well as a large fundamental gap $E_g\equiv E_0-\Delta_2$, see Section~\ref{sec:effham} and Table~\ref{tab:bandparameters}.  In this case the energies of bands entering the energy denominators in Eq.~\eqref{eq:mformula} are given by
\begin{align}
  \e_{c}&= E_0 +\frac{\hbar^2 k_\perp^2}{2m_\perp^{c}}+ \frac{\hbar^2 k_z^2}{2m_\parallel^{c}},\nonumber\\
  \e_{v}^\pm&=\frac{\hbar^2 k_\perp^2}{2m_\perp^{v}}+ \frac{\hbar^2 k_z^2}{2m_\parallel^{v}}\pm\sqrt{\D_2^2+\beta_v k_z^2},\nonumber\\
 \e_{h}&= -2\Delta_1+\frac{\hbar^2 k_\perp^2}{2m_\perp^{h}}+ \frac{\hbar^2 k_z^2}{2m_\parallel^{h}}.
\end{align}
 We reiterate that all mass parameters already include corrections due to the band interaction.

Further, the matrix elements of the transverse velocity operator, $\p_{x,y} h_\kk$, which are off-diagonal in $c,v,h$-space, are linear in the interband couplings, $P_i$. Therefore, to lowest ($O(P_1^2,P_4^2)$) order  one can use the unperturbed band wave functions to calculate the $z$-component of the orbital magnetic moment from Eq.~\eqref{eq:mformula}. Given our neglect of the conduction and hole band spin splittings, the simplest choice of the unperturbed basis is
\begin{align}\label{eq:unperturbedus}
  |u_c^{+(-)}\ket&=(1(0),0(1),0,0,0,0)^T,\nonumber\\
  |u_v^{+(-)}\ket&=(0,0,\pm\sqrt{\frac{1\pm\gamma_{k_z}}{2}},\sqrt{\frac{1\mp\gamma_{k_z}}{2}},0,0)^T,\nonumber\\
   &\gamma_{k_z}=\frac{\beta_v k_z}{\sqrt{\Delta_2^2+\beta_v^2k_z^2}}\nonumber\\
  |u_h^{+(-)}\ket&=(0,0,0,0,1(0),0(1))^T,\nonumber\\
\end{align}
Note that the velocity operator $\p_\kk h _\kk$ does not have matrix elements between the upper and lower valence bands, hence $\e_v^-$ and $|u_v^{-}\ket$ are given for completeness.

The final expression for the $z$-component of the orbital magnetic moment in the valence band, $m^{\textrm {orb}}_z \equiv\bm m^{\textrm{orb}}_v\cdot\bm z$, becomes
\begin{align}
  m^{\textrm{orb}}_z (\kk)=\frac{e}{\hbar}\left[\frac{P_1^2}{\e^+_v-\e_c}+\frac{P_4^2}{\e^+_v-\e_h}\right]\frac{\beta_v k_z}{\sqrt{\Delta_2^2+\beta_v^2k_z^2}}.
\end{align}

The spin contribution to the magnetic moment of the upper valence band can be evaluated by directly calculating the expectation of the spin operator in that band, see Eq.~\eqref{eq:magneticmoment}. Direct calculation yields the spin magnetic moment $\bm m^{\textrm{spin}}_v=m^{\textrm{spin}}_z\bm z$, with
\begin{align}
  m^{\textrm{spin}}_z (\kk)=\chi\frac{e\hbar}{m_0}\frac{\beta_v k_z}{\sqrt{\Delta_2^2+\beta_v^2k_z^2}},
\end{align}
where $\chi\approx 1/2$.

Finally, the expression for the $z$-component of the intrinsic orbital magnetic moment in the upper valence band, $m^{\textrm{int}}_z$, is given by
\begin{align}\label{eq:totalintrinsicm}
  m^{\textrm{int}}_z (\kk)=\frac{e}{\hbar}\left[\frac{P_1^2}{\e^+_v-\e_c}+\frac{P_4^2}{\e^+_v-\e_h}+\chi\frac{\hbar^2}{m_0}\right]\frac{\beta_v k_z}{\sqrt{\Delta_2^2+\beta_v^2k_z^2}}.
\end{align}
This expression allows to estimate the relative size of orbital and spin contributions to the intrinsic magnetic moment near the H-point:
\begin{align}
  \frac{|m^{\textrm{orb}}|}{|m^{\textrm{spin}}|}=\left|\frac{P_1^2}{E_0-\Delta_2}-\frac{P_4^2}{2\Delta_1+\Delta_2}\right| \frac{m_0}{\chi \hbar^2}\approx 3.3.
\end{align}

We note that an expression similar to Eq.~\eqref{eq:totalintrinsicm} for the intrinsic \textit{angular momentum} of quasiparticles near the H-point was obtained in Ref.~\onlinecite{Farbshtein2012}.

\subsubsection{Extrinsic magnetic moment: skew scattering contribution}

The skew-scattering contribution to the effective magnetic moment, Eq.~\eqref{eq:skmomentresults}, involves the antisymmetric part of the scattering rate, given by Eq.~\eqref{eq:antisymmetricWPBphase}. The skew scattering rate involves Fourier transforms of the impurity potential, as well as the overlaps of the periodic parts of the Bloch functions. First, we address the Fourier transform of the scattering potential by noting that the weak dependence of the mobility and scattering time on the dopant density (antimony) in tellurium\cite{Bresler1970} suggests that the electron scattering at low temperatures mostly happens on background neutral defects. Assuming these are characterized by a potential that can be considered short range for the typical values of the Fermi wave lengths of holes in tellurium (10-100 nm), from now on we will set
\begin{align}
V^{imp}(\qq)\approx V^{imp}(0)\equiv V_0.
\end{align}
For continuity of presentation, we defer further numerical estimates for $V_0$ until Section~\ref{sec:KME}.

The simplifications described above lead to the following expression for the antisymmetric part of the scattering rate:
\begin{align}\label{eq:asymWsimplified}
  W^{\textrm{A}}_{\kk\kk'}\approx&-\frac{(2\pi)^2}{\hbar}n_i V_0^3\int (d\kk'')\d(\e_\kk-\e_{\kk''})\d(\e_{\kk'}-\e_{\kk''})\nonumber\\
  &\times {\rm Im}\left[\bra u_\kk|u_{\kk'}\ket \bra u_{\kk'}|u_{\kk''}\ket\bra u_{\kk''}|u_\kk\ket\right].
\end{align}

In Appendix~\ref{sec:appendixB} we evaluate the scattering rate~\eqref{eq:asymWsimplified} for the valence band, and obtain
\begin{align}\label{eq:skewWfinal_maintext}
  W^{\textrm{A}}_{\kk\kk'}=&F_{\kk\kk'}\bm z\cdot\kk\times\kk'\d(\e_{v\kk}-\e_{v\kk'}),
\end{align}
where the symmetric prefactor $F_{\kk,\kk'}=F_{\kk',\kk}$ is given in Eq.~\eqref{eq:prefactorF}. Expression~\eqref{eq:skewWfinal_maintext} is dictated by the symmetry group of Te ($D_3$). Its physical meaning can be understood by considering the generalization of the usual Mott scattering rate, $W^{\textrm{A}}_{\textrm{Mott}}\propto \bm s \cdot\kk\times\kk'\d(\e_{\kk}-\e_{\kk'})$ , on a spin-orbit coupled atom in free space to the case of the spin-split valence band of tellurium. In this case the spin, $\bm s$, of the scattered particle is substituted with the properly symmetrized effective orbital moment acquired by a charge carrier due to its motion along the helices of Te atoms in $z$-direction, $\bm s\to (k_z+k_{z'})\bm z$, see Ref.~\onlinecite{Rou2017} for more details.

The expression for the skew-scattering contribution to the effective orbital magnetic moment is obtained using Eq.~\eqref{eq:skmomentresults}. For its $z$-component, which determines the KME for current flowing along the optical axis, we obtain
\begin{align}\label{eq:mskfinal_maintext}
  m^{\textrm{sk}}_z=\frac{e\hbar^2\tau^2}{m_\perp^2}\bm \int (d\kk')W^{\textrm{A}}_{\kk\kk'}\bm z\cdot\kk\times\kk'.
  \end{align}
The full expression is given in Appendix~\ref{sec:appendixB}, see Eq.~\eqref{eq:mskfinal}. One can make a general remark regarding the dependence of  $m^{\textrm{sk}}_z$ on the impurity density: as $\tau\propto 1/n_i$ and $W^{\rm A}\propto n_i$, $m^{\textrm{sk}}_{z}$ is inversely proportional to the impurity density, or, equivalently, proportional to the mean free path of charge carriers. That is, we expect the skew-scattering contribution to dominate in sufficiently clean  samples.

\subsubsection{Extrinsic magnetic moment: side jump contribution}

The central quantity determining the side-jump contribution to the magnetic moment, Eq.~\eqref{eq:sjmomentresults}, is the side jump displacement $\d \rr_{\kk\kk'}$, Eq.~\eqref{eq:sidejumpPBphase}. The calculation of this quantity can be found in Appendix~\ref{sec:appendixC}, while here we present the general form of the result:
\begin{align}\label{eq:deltar_maintext}
  \d\rr_{\kk\kk'}=-A_{\kk}\gamma_{k_z}\bm z\times \kk+A_{\kk'}\gamma_{k_z'}\bm z\times \kk'-B_{\kk\kk'}\bm z\times (\kk-\kk').
\end{align}
The coefficients $A_{\kk}$ and $B_{\kk\kk'}=B_{\kk'\kk}$ can be found in Eq.~\eqref{eq:sidejump_appendix}; $\gamma_{k_z}$ was introduced in Eq.~\eqref{eq:unperturbedus}.

The side-jump contribution to the effective magnetic moment is given by Eq.~\eqref{eq:sjmomentresults}. Due to the $\kk'$ integration in Eq.~\eqref{eq:sjmomentresults}, only the terms that are proportional to $\bm z\times \kk$ in Eq.~\eqref{eq:deltar_maintext} contribute to the magnetic moment, the $z$-component of which can be written as
\begin{align}\label{eq:msj_maintext}
  m^{\textrm{sj}}_{z}=e\tau\int (d\kk')W^{\textrm{S}}_{\kk\kk'} (A_{\kk}\gamma_{k_z}+B_{\kk\kk'})(\bm z\times \kk)\cdot(\bm z\times\vv_\kk).
\end{align}

The full expression is given in Appendix~\ref{sec:appendixC}, see Eq.~\eqref{eq:msj_appendix}. As is common for phenomena related to the side jump, the dependence of $m^{\textrm{sj}}_{z}$ on the impurity density mimics that of the intrinsic contribution. Indeed, since $\tau\propto 1/n_i$ and $W^{\rm S}\propto n_i$, $m^{\textrm{sj}}_{z}$ is independent of the impurity density.

\subsection{Kinetic magnetoelectric effect}\label{sec:KME}

In this Section, we use Eqs.~\eqref{eq:totalintrinsicm}, \eqref{eq:mskfinal_maintext}, and \eqref{eq:msj_maintext} to calculate the KME in tellurium. We consider the magnetization response to a transport current along the optical ($z$) axis of a tellurium crystal. In practice, it is useful to specify the response of magnetization directly to the transport current,
\begin{align}\label{eq:alphazz}
  M_z=\alpha_{zz}j_{z},
\end{align}
rather than to the electric field driving it.  In the rest of this Section, we calculate the KME response coefficient $\alpha_{zz}$.

Assuming the current flow is caused by electric field $\EE=(0,0,E_z)$, and restricting ourselves to the case of low temperature, we can write the correction to the distribution of electrons in the upper valence band in the relaxation time approximation as
\begin{align}\label{eq:deltaf}
  \d f_{v}(\kk)=\tau(\mu)eE_z v_z(\kk)\d(\e_v^+(\kk)-\mu).
\end{align}
Here $\mu$ is the chemical potential, and $\tau(\mu)$ is the transport relaxation time; the standard $\d$-function factor comes from the energy derivative of the equilibrium distribution function at low temperatures.

The expression for the conductivity of the crystal is then given by
\begin{align}
  \sigma_{zz}(\mu)=e^2 \tau(\mu)\int(d\kk) v_z^2(\kk)\d(\e_v^+(\kk)-\mu).
\end{align}
Using the conductivity and expression for the current-induced magnetization~\eqref{eq:KME}, the coefficient $\a_{zz}$  can be written as
\begin{align}\label{eq:alpha_definition}
  \a_{zz}(\mu)=\frac{ \int(d\kk)
 m_z(\kk)v_z(\kk)\d(\e_v^+(\kk)-\mu)}{e\int(d\kk) v_z^2(\kk)\d(\e_v^+(\kk)-\mu)}.
\end{align}

It is apparent from either Eq.~\eqref{eq:alpha_definition}, or Eq.~\eqref{eq:alphazz}, that $\alpha_{zz}$ is measured in units of length. Further, it is also clear from Eq.~\eqref{eq:alpha_definition} that in contrast to the skew-scattering contribution, the intrinsic and side-jump contributions to the KME response coefficient are relatively insensitive to a particular disorder model, while the skew scattering contribution does require a choice of the latter. To pick a disorder model, we note that experimentally the low-temperature mobility of tellurium is known to depend rather weakly on the carrier density: it roughly drops from $4500\, \textrm{cm}^2/\textrm{V}\cdot \textrm {s}$ to $3100\, \textrm{cm}^2/\textrm{V}\cdot \textrm {s}$ upon varying the hole density from $2\cdot 10^{17} \textrm{cm}^{-3}$ to $2\cdot 10^{18} \textrm{cm}^{-3}$ (see Ref.~\onlinecite{Bresler1970} for details). Within the relaxation time approximation defined by Eq.~\eqref{eq:deltaf}, this corresponds to a decrease in the scattering time from $0.8\,\textrm{ ps}$ at low density to $0.3\,\textrm{ ps}$ at the highest density. Given the semiphenomenological nature of our treatment of disorder scattering, it is justified to consider $\tau(\mu)$ as a constant, $\tau(\mu)\sim 0.5\, \textrm{ps}$.

Further, the skew-scattering contribution explicitly depends on the strength of the impurity potential, $V_0$. This quantity can be estimated from the standard expression for the scattering time:
\begin{align}\label{eq:1overtau}
  \frac{1}{\tau(\mu)}\sim\frac{2\pi}{\hbar}n_i \nu(\mu) V_0^2.
\end{align}
For typical defect densities\cite{regel2001defect}, $n_i\sim 10^{17}-10^{18}\,\textrm{cm}^{-3}$, and $\tau(\mu)\sim 0.5\cdot 10^{-12}\,\textrm{s}$ for $\mu\sim 0$ (midpoint between the two valence bands), we obtain $V_0\sim 5\,\textrm{eV}\cdot \textrm{nm}^3$. This estimate compares favorably with the scale of $V_0$ for a hydrogen-like impurity with an effective Bohr radius $a_B$:
\begin{align}\label{eq:V0estimate}
  V_0\sim\frac{e^2}{4\pi\varepsilon_0\varepsilon a_B}a_B^3\sim 10\, \textrm{eV}\cdot\textrm{nm}^3, \,\,a_B=\frac{4\pi\varepsilon_0\varepsilon \hbar^2}{m^* e^2},
\end{align}
where we took $\varepsilon\approx 34$ as the effective isotropic dielectric constant of tellurium, and $m^*\sim 0.1 m_0$ is the typical effective hole mass. In what follows, we will take the estimate~\eqref{eq:V0estimate} to calculate the extrinsic contributions, and will eliminate the unknown impurity concentration by means of Eq.~\eqref{eq:1overtau}, setting where appropriate
\begin{align}
  n_iV_0^2\sim \frac{\hbar}{2\pi \tau(\mu)\nu(\mu)}.
\end{align}
\begin{figure}
  \centering
  \includegraphics[width=3in]{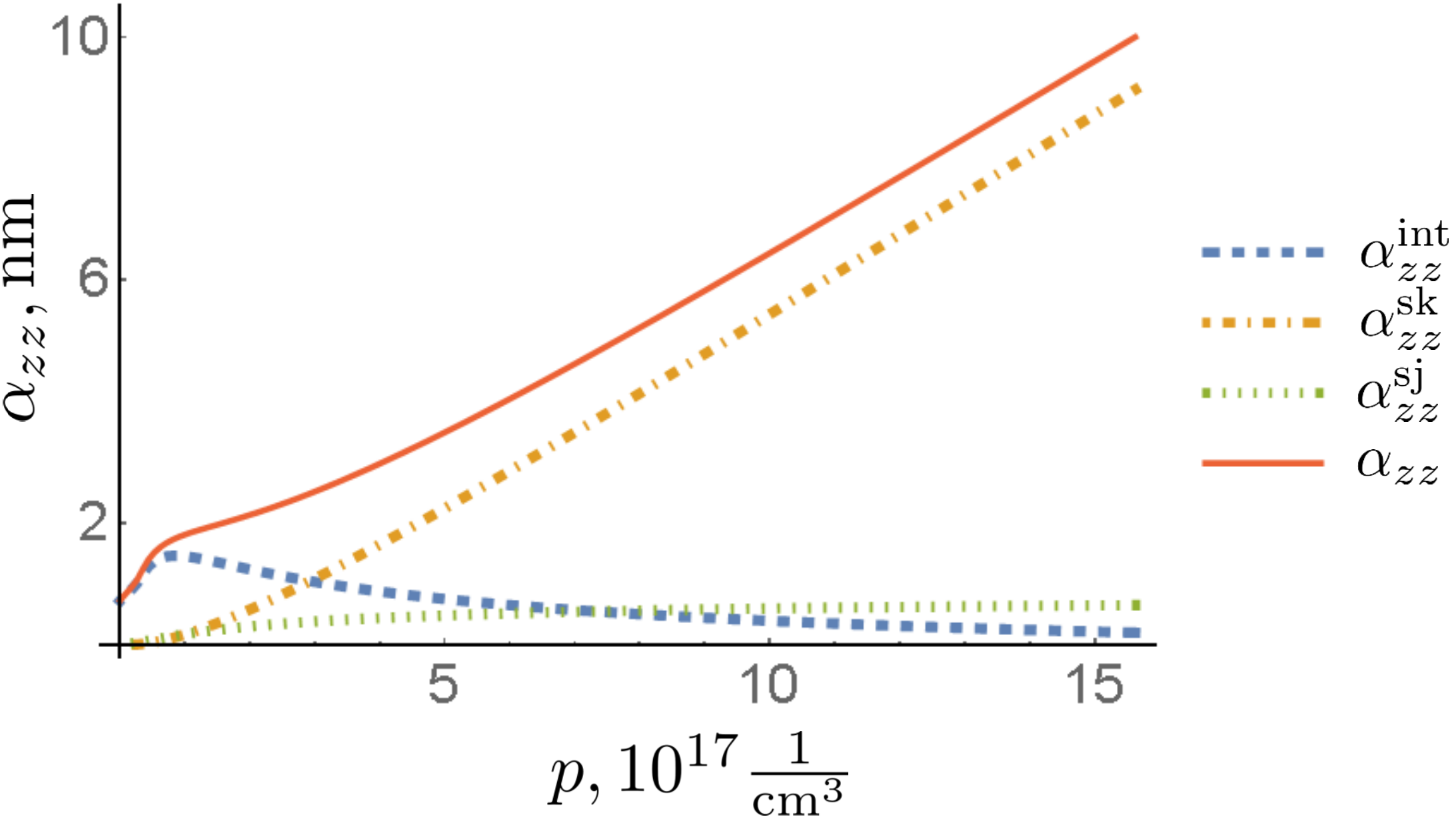}
  \caption{(Color online) The dependence of contributions to the KME response coefficient, $\alpha_{zz}$ (red solid line), on the hole density in tellurium. Blue dashed, orange dot-dashed, and green dotted lines depict the individual contributions to  $\alpha_{zz}$ from intrinsic,  $\alpha_{zz}^{\textrm{int}}$, and extrinsic, $\alpha_{zz}^{\textrm{sk/sj}}$, mechanisms. }\label{fig:alpha}
\end{figure}

Combining Eqs.~\eqref{eq:totalintrinsicm}, \eqref{eq:mskfinal_maintext}, and \eqref{eq:msj_maintext} with the expression for the KME response coefficient, Eq.~\eqref{eq:alpha_definition}, we obtain the intrinsic ($\alpha_{zz}^{\textrm{int}}$), skew-scattering ($\alpha_{zz}^{\textrm{sk}}$), and side-jump ($\alpha_{zz}^{\textrm{sj}}$)  contributions the KME, which originate from the corresponding magnetic moments. The result must be multiplied by a factor of 2 to account for the identical contribution from the $H'$ point. The result of numerical evaluation of the three contributions for different hole densities is presented in Fig.~\ref{fig:alpha}, and is one of the main results of this paper.

The notable features of Fig.~\ref{fig:alpha} are \textit{i}) the non-monotonic behavior of the intrinsic contribution, and \textit{ii}) the competition between different mechanisms: the intrinsic contribution is the leading one at low densities, while the skew scattering one wins at a large enough level of hole doping. These observations can be explained using the defining expressions, Eqs.~\eqref{eq:totalintrinsicm}, \eqref{eq:mskfinal_maintext}, \eqref{eq:msj_maintext} and~\eqref{eq:alpha_definition}, see below.

The discussion of features seen in Fig.~\ref{fig:alpha} is greatly facilitated by introducing a dimensionless parameter $\eta$ given by
\begin{align}
  \eta=\frac{\hbar^2\Delta_2}{|m_z^v|\beta_v^2}\approx 0.76.
\end{align}
Note that $m_z^v$ is the effective mass in the $z$-direction, not to be confused with the $z$-component of the magnetic moment.

First, we address the shape of the intrinsic contribution curve. We note that for sufficiently small doping, when the chemical potential $\mu$ (counted from the center of mass of the upper valence bands) satisfies
\begin{align}
  \frac{1}{2}\left(\eta+\frac{1}{\eta}\right)\Delta_2>\mu>\Delta_2,
\end{align}
there are two disjoint Fermi surfaces. For very low doping, the two Fermi surfaces are centered around the valence band ``humps'', whose maximaare  located on the H-K line at
\begin{align}
  k_0=\sqrt{\frac{1}{\eta^2}-1}\frac{\Delta_2}{\beta_v}\approx 0.02 \frac{1}{\textrm{\AA}},
\end{align}
as seen in Fig.~\ref{fig:dispersions}. Obviously, $v_z(k_z=k_0)=0$, and near $k_z=k_0$ both $v_z$ and $m_z$ admit the following Taylor expansions:
\begin{align}
  v_z(\kk)&\approx -(1-\eta^2)\frac{\hbar (k_z-k_0)}{|m_z^v|},\nonumber\\
  m_z^{\textrm{int}}(\kk)&\approx m_z^{\textrm{int}}(k_0)+\frac{\partial m_z^{\textrm{int}}(k_0)}{\partial k_0}(k_z-k_0).
\end{align}
It is then obvious that the finite value of  $\alpha_{zz}^{\textrm{int}}$ at vanishing hole density is given by
\begin{align}
  \alpha_{zz}^{\textrm{int}}=-\frac{2}{\eta (1-\eta^2)}\frac{\hbar \Delta_2}{e\beta_v^2}\frac{\partial m_z^{\textrm{int}}(k_0)}{\partial k_0}\approx 1\textrm{nm},
\end{align}
where the overall factor of 2 counts the contributions from H-  and H$'$-points. In other words, the integrals in the numerator and denominator of Eq.~\eqref{eq:alpha_definition} are both small, but their ratio is finite. The upward trend at small doping is due to the fact that $m_z^{\textrm{int}}$ is a sign-definite function at small $k_z$, while $v_z(\kk)$ changes its sign from positive to negative, which makes the numerator of Eq.~\eqref{eq:alpha_definition} grow faster than its denominator.  At large level of doping, for which the Fermi wave vector satisfies $k_F\gg\Delta_2/\beta_v$, $m_z^{\textrm{int}}$ almost saturates, but $|v_z(\kk)|$ keeps growing with $k_z$, hence the decrease in $\alpha_{zz}^{\textrm{int}}$. We note that the existence of the maximum in the dependence of $\alpha_{zz}^{\textrm{int}}$ on hole density is a consequence of the double-hump dispersion of tellurium, which in turn comes from $\eta<1$. One can imagine a crystal with the same crystal structure as tellurium, but with parameters that correspond to $\eta>1$, for which there will be no maximum in $\alpha_{zz}^{\textrm{int}}$.

It is apparent from Fig.~\ref{fig:alpha} that the skew-scattering and side-jump contributions are strongly suppressed at small densities. This is due to the fact that in order to have side jump and skew scattering processes, an electron must have a finite momentum component, $k_\perp$, in the basal plane. At small densities, the Fermi surfaces are small pockets around $\kk=(0,0,\pm k_0)$, for which $k_\perp\ll k_z$, and the extrinsic contributions are strongly suppressed. The skew-scattering contribution contains $k_\perp^4$, see Eqs.~\eqref{eq:skewWfinal_maintext} and~\eqref{eq:mskfinal_maintext}, while the side-jump one is suppressed as $k_\perp^2$, see Eq.~\eqref{eq:msj_maintext}. This means that the skew-scattering contribution sees the strongest suppression at small density, and the most rapid growth as the Fermi surface expands. Overall, this leads to the picture apparent from Fig.~\ref{fig:alpha}: the intrinsic contribution is the leading contribution at small densities, which skew scattering overpowers it at a certain finite density. The side jump contribution is always subleading.

\section{Discussion}\label{sec:conclusions}

We have considered the kinetic magnetoelectric effect in a typical helical crystal - p-doped trigonal tellurium. The KME is identical to the phenomenon of current-induced magnetization, with both orbital and spin contributions to magnetization taken into account. The KME has contributions from quasiparticle magnetic moments of intrinsic (band structure) and extrinsic (skew and side-jump scattering off impurities) origin. In tellurium, we showed that it is the intrinsic contribution to magnetization that dominates the KME response for hole densities of about $(1-3)\cdot 10^{17}\textrm{cm}^{-3}$. For larger densities, the extrinsic mechanism related to impurity skew scattering overpowers the intrinsic one, while the side-jump related contribution is never dominant, or even considerable. These findings are summarized in Fig.~\ref{fig:alpha}.

The magnitude of the KME in tellurium is quantified by the corresponding response coefficient $\alpha_{zz}$, see Eq.~\eqref{eq:alphazz}. The typical value for tellurium is $\alpha_{zz}\sim 5 \textrm{nm}$. For a transport current of $j_z\sim 10^3 \textrm{A}/\textrm{cm}^2$, this corresponds to a magnetization of $\mu_0M_z\sim 60 \textrm{nT}$. To appreciate this figure, one can calculate the equivalent spin polarization of free carriers with the same density, $p\sim 5\cdot 10^{17}\textrm{cm}^{-3}$. Such spin polarization is about 1\%, consistent with findings of Ref.~\onlinecite{Farbshtein2012}, which is two orders of magnitude larger than the one attained in experiments on strained semiconductors~\cite{Kato2004magnetization,Kato2005magnetization}.

Alternatively, one can ask what external magnetic field is needed to induce a magnetization of $60\,\textrm{nT}$, given the magnetic susceptibility of tellurium, $\mu_0\chi_m=3\cdot 10^{-7}$ (see Ref.~\onlinecite{Bresler1972} and references therein). Such field is about $0.2\, \textrm{T}$. This exceeds the stray field of a current with current density $j_z\sim 10^3 \textrm{A}/\textrm{cm}^2$ in a sample whose transverse dimensions are  up to $10\,\textrm{cm}$. Thus the KME is a good way to induce magnetic moments without the drawback of having large stray magnetic fields. This can have broad applications in spin electronics.

The possibile spintronics applications offered by helical metals can be further evaluated by comparing the present case of the KME in tellurium to spin-orbit induced magnetic torques in magnetic semiconductors. The maximum induced magnetization in tellurium corresponds to the chemical potential located close to the top of the lower valence band, $\mu=-\Delta_2$: for higher hole dopings ($\mu<-\Delta_2$) the KME in the two valence bands start to partially cancel each other. At $\mu=-\Delta_2$, the hole density is $6\cdot10^{18}\textrm{cm}^{-3}$, and the KME coefficient reaches its maximum value $\alpha_{zz}\sim 40\textrm{nm}$, or $\mu_0\alpha_{zz}\sim 0.5\cdot 10^{-9}\textrm{T}\, \textrm{cm}^{2} \,\textrm{A}^{-1}$. This is on par with the figure reported\cite{Chernyshov2009} for strained (Ga,Mn)As ferromagnetic semiconductor at hole density of $2\cdot10^{20}\textrm{cm}^{-3}$ - almost two orders of magnitude larger than in tellurium. The ability of tellurium to efficiently convert transport current into magnetization results from the strongest possible breaking of inversion symmetry, broken by the atomic-scale helices, and tellurium's strong spin-orbit coupling.

\textit{Note added:} When this work was nearing completion, we became aware of a complementary first-principle study of gyrotropic effects in tellurium in Ref.~\onlinecite{tsirkin2017}, which focused on the intrinsic effects. The magnitude of the intrinsic KME reported in Ref.~\onlinecite{tsirkin2017} can be brought in agreement with the one obtained in the present work upon some adjustment of overall numerical factors in the former.
\acknowledgments
This work was supported by the National Science Foundation Grant No. DMR-1409089.

\appendix

\onecolumngrid

\section{Bloch state overlaps and the Pancharatnam phase in the valence band of Te}\label{sec:appendixA}

As is seen from Eqs.~\eqref{eq:antisymmetricWPBphase} and \eqref{eq:sidejumpPBP} of the main text, the main two quantities determining the skew scattering and side jump contributions to the effective magnetic moment of quasiparticles are
\begin{align}
\Phi_{\kk,\kk',\kk''}&={\rm arg}\left[\bra u_{v\kk}|u_{v\kk'}\ket \bra u_{v\kk'}|u_{v\kk''}\ket\bra u_{v\kk''}|u_{v\kk}\ket\right],\nonumber\\
  \Upsilon_{\kk,\kk',\kk''}&={\rm Im}\left[\bra u_{v\kk}|u_{v\kk'}\ket \bra u_{v\kk'}|u_{v\kk''}\ket\bra u_{v\kk''}|u_{v\kk}\ket\right].
\end{align}
In these expressions we suppressed the superscript $``+"$ on $|u_{v\kk}\rangle$, as it always refers to the topmost valence band. In this Appendix we illustrate the calculation of $\Phi_{\kk,\kk',\kk''}$ and $\Upsilon_{\kk,\kk',\kk''}$ to the leading order in the band interaction.

First, we note that overlaps of the periodic parts of the Bloch functions in the topmost valence band calculated using the unperturbed states, Eqs.~\eqref{eq:unperturbedus},  are real. Therefore, to calculate $\Upsilon_{\kk,\kk',\kk''}$ and $\Phi_{\kk,\kk',\kk''}$ one must take into account band interaction. To the leading order, $\textrm{Im}\bra u_{v\kk}|u_{v\kk'}\ket\sim \textrm{O}(P_1^2,P_4^2)$, and if we restrict the calculation of $\Upsilon_{\kk,\kk',\kk''}$ and $\Phi_{\kk,\kk',\kk''}$ to the same order we obtain
\begin{align}\label{eq:upsilonperturb}
 \Upsilon_{\kk,\kk',\kk''}&={\rm Im}\left[\bra u_{v\kk}|u_{v\kk'}\ket \right]\Gamma^+_{k_z',k_z''}\Gamma^+_{k_z'',kz}
 +{\rm Im}\left[\bra u_{v\kk'}|u_{v\kk''}\ket\right]\Gamma^+_{k_z,k_z'}\Gamma^+_{k_z'',k_z}
 +{\rm Im}\left[\bra u_{v\kk''}|u_{v\kk}\ket\right]\Gamma^+_{k_z,k_z'}\Gamma^+_{k_z',k_z''}, \\
 \Gamma^\pm_{k_z,k_z'}&=\frac12 \left(\sqrt{(1+\gamma_{k_z})(1+\gamma_{k_z'})}\pm \sqrt{(1-\gamma_{k_z})(1-\gamma_{k_z'})}\right).\nonumber
\end{align}
The function $\Gamma^+_{k_z,k_z'}$ is the overlap of the unperturbed wave functions of the upper valence band, Eq.~\eqref{eq:unperturbedus}, at wave vectors $\kk$ and $\kk'$; $\Gamma^-_{k_z,k_z'}$ has been introduced for future convenience.

By the same token, we obtain for $\Phi_{\kk,\kk',\kk''}$:
\begin{equation}\label{eq:phiperturb}
  \Phi_{\kk,\kk',\kk''}=\frac{{\rm Im}\left[\bra u_{v\kk}|u_{v\kk'}\ket \right]}{\Gamma^+_{k_z,k_z'}}
  +\frac{{\rm Im}\left[\bra u_{v\kk'}|u_{v\kk''}\ket \right]}{\Gamma^+_{k_z',k_z''}}
  +\frac{{\rm Im}\left[\bra u_{v\kk''}|u_{v\kk}\ket \right]}{\Gamma^+_{k_z'',k_z}}.
\end{equation}

From Eqs.~\eqref{eq:upsilonperturb} and \eqref{eq:phiperturb} it is apparent that the key quantity that determines the geometric phases in the electronic band structure is ${\rm Im}\left[\bra u_{v\kk}|u_{v\kk'}\ket \right]$. In what follows, we turn to the evaluation of such terms. In doing so, we restrict ourselves to the interaction between the valence and conduction bands. The additive contribution from the hole bands can be obtained in exactly the same manner \textit{mutatis mutandis}.

In the subspace spanned by the conduction and valence bands, the effective Hamiltonian \eqref{eq:3x3hamiltonian} (or \eqref{eq:matrixham}) can be represented as
\begin{align}
  h_\kk=h_\kk^0+\delta h_\kk,
\end{align}
with
\begin{align}
h_\kk^0=\left(
\begin{array}{cc}
h_{cc}& 0  \\
0& h_{vv}  \\
\end{array}
\right),\,\,
\d h_\kk=\left(
\begin{array}{cc}
0& h_{cv}  \\
h_{vc}& 0 \\
\end{array}\right).
\end{align}
Here $\d h_\kk$ describes the band interaction, and the corresponding perturbative correction to the topmost valence band wave function is
\begin{align}
  |\delta u_{v}\ket=\sum_{\zeta=\pm}|u^{\zeta}_c\ket\frac{\bra u_c^{\zeta}|\d h_\kk|u_v\ket}{\e_{v}^+-\e_{c}^\zeta}.
\end{align}
As we neglect the spin splitting in the conduction band, $\e_{c}^\pm=\e_c$, the summation over $\zeta$ can be performed to yield
\begin{align}\label{ap:perturbation}
  |\delta u_{v\kk}\ket=\frac{P_c\d h_\kk|u_{v\kk}\ket}{\e_{v}^+(\kk)-\e_{c}(\kk)},
\end{align}
where $P_c=\textrm{diag}(1,1,0,0)$ is the projector onto the conduction band, and we reinstated the momentum index for future convenience.

It is easy to show that the first order perturbation theory for $|u_{v\kk}\ket$ -- Eq.~\eqref{ap:perturbation} -- is sufficient to calculate $\bra u_{v\kk}|u_{v\kk'}\ket$ to second order in band interaction parameters. Using Eq.~\eqref{ap:perturbation} we obtain
\begin{align}
  \bra u_{v\kk}|u_{v\kk'}\ket&=\Gamma^+_{k_z,k_z'}+\frac{\bra u_{v\kk}|\d h_\kk P_c\d h_{\kk'}|u_{v\kk'}\ket}{[\e_{v}^+(\kk)-\e_{c}(\kk)][\e_{v}^+(\kk')-\e_{c}(\kk')]}.
\end{align}
The first term on the right hand side of this expression comes from the unperturbed wave functions, Eqs.~\eqref{eq:unperturbedus}. Using
\begin{align}
\d h_\kk P_c\d h_{\kk'}=
\left(
\begin{array}{cc}
0& 0 \\
0& h_{vc}(\kk) h_{cv}(\kk')\\
\end{array}
\right),
\end{align}
and the explicit form of matrices $h_{cv}$ and $h_{vc}$, Eq.~\eqref{eq:matrixham},
we obtain
\begin{align}\label{eq:overlap}
  \textrm{Im} \bra u_{v\kk}|u_{v\kk'}\ket&=
\frac{P_1^2\Gamma^-_{k_z,k_z'}}
  {[\e_{v}^+(\kk)-\e_{c}(\kk)][\e_{v}^+(\kk')-\e_{c}(\kk')]}\,\bm z\cdot\kk\times\kk'.
\end{align}
The contribution from the hole band is an additive term with trivial replacements: $P_1\to P_4$, $\e_c\to \e_h$. Equations~\eqref{eq:upsilonperturb}, \eqref{eq:phiperturb}, and \eqref{eq:overlap} form the basis for the calculation of the skew scattering and side jump contributions to the effective magnetic moment, performed in Appendix~\ref{sec:appendixB}.

\section{Calculation of the skew scattering contributions to the effective magnetic moment}\label{sec:appendixB}

To calculate the skew scattering contribution to the magnetic moment, we first address the skew scattering rate,
\begin{align}
  W^{\textrm{A}}_{\kk\kk'}=&-\frac{(2\pi)^2}{\hbar}n_i V_0^3\d(\e_{v\kk}-\e_{v\kk'})\int (d\kk'')\d(\e_{v\kk'}-\e_{v\kk''})\Upsilon_{\kk,\kk',\kk''}.
\end{align}
Because of the angular part of the $\kk''$ integral, only the first term in right hand side of Eq.~\eqref{eq:upsilonperturb} contributes to $W^\textrm{A}_{\kk\kk'}$. Taking into account both the conduction and the hole bands contributions, we obtain
\begin{align}\label{eq:skewWfinal}
  W^{\textrm{A}}_{\kk\kk'}=&F_{\kk,\kk'}\bm z\cdot\kk\times\kk'\d(\e_{v\kk}-\e_{v\kk'}),
\end{align}
where
\begin{align}\label{eq:prefactorF}
  F_{\kk,\kk'}=&-\frac{(2\pi)^2}{\hbar}n_i V_0^3\int (d\kk'')\d(\e_{v\kk'}-\e_{v\kk''})\left[
\frac{P_1^2\Gamma^-_{k_z,k_z'}\Gamma^+_{k_z',k_z''}\Gamma^+_{k_z'',kz}}
  {[\e_{v}^+(\kk)-\e_{c}(\kk)][\e_{v}^+(\kk')-\e_{c}(\kk')]}+\frac{P_4^2\Gamma^-_{k_z,k_z'}\Gamma^+_{k_z',k_z''}\Gamma^+_{k_z'',kz}}
  {[\e_{v}^+(\kk)-\e_{h}(\kk)][\e_{v}^+(\kk')-\e_{h}(\kk')]}\right].
\end{align}
Finally, using Eq.~\eqref{eq:skmomentresults} of the main text, we obtain the $z$-component of the skew-scattering magnetic moment:
  \begin{align}
  m^{\textrm{sk}}_z&=-2\pi^2\frac{e\hbar\tau^2}{(m^v_\perp)^2}n_i V_0^3\int (d\kk') (d\kk'')(k_\perp {k_\perp'})^2 \d(\e_{v\kk}-\e_{v\kk'})
  \d(\e_{v\kk}-\e_{v\kk''})\Gamma^-_{k_z,k_z'}\Gamma^+_{k_z',k_z''}\Gamma^+_{k_z'',kz}\nonumber\\
  &\times\left[\frac{P_1^2}
{[\e_{v}^+(\kk)-\e_{c}(\kk)][\e_{v}^+(\kk')-\e_{c}(\kk')]}+\frac{P_4^2}
  {[\e_{v}^+(\kk)-\e_{h}(\kk)][\e_{v}^+(\kk')-\e_{h}(\kk')]}\right].
  \end{align}

  \section{Calculation of the side jump contributions to the effective magnetic moment}\label{sec:appendixC}

To calculate the side jump contribution to the magnetic moment, we first address the side jump displacement itself,
\begin{equation}
  \d\rr_{\kk\kk'}=\lim_{\kk''\to\kk}\p_{\kk''}\Phi_{\kk,\kk',\kk''}+\lim_{\kk''\to\kk'}\p_{\kk''}\Phi_{\kk,\kk',\kk''}.
\end{equation}
It is clear that only the last two terms on the right hand side of Eq.~\eqref{eq:phiperturb} contribute to $\d \rr_{\kk\kk'}$, and using Eq.~\eqref{eq:overlap} we obtain
\begin{align}\label{eq:sidejump_appendix}
  \d\rr_{\kk\kk'}&=-\left(\frac{P_1^2}{[\e_{v}^+(\kk)-\e_{c}(\kk)]^2}+\frac{P_4^2}{[\e_{v}^+(\kk)-\e_{h}(\kk)]^2}\right)\,\gamma_{k_z}\bm z\times\kk
+\left(\frac{P_1^2}{[\e_{v}^+(\kk')-\e_{c}(\kk')]^2}+\frac{P_4^2}{[\e_{v}^+(\kk')-\e_{h}(\kk')]^2}\right)\,\gamma_{k_z'}\bm z\times\kk'\nonumber\\
  &-\left(
\frac{P_1^2}{[\e_{v}^+(\kk')-\e_{c}(\kk')][\e_{v}^+(\kk)-\e_{c}(\kk)]}+
\frac{P_4^2}{[\e_{v}^+(\kk')-\e_{h}(\kk')][\e_{v}^+(\kk)-\e_{h}(\kk)]}\right)\,\frac{\Gamma^-_{k_z',k_z}}{\Gamma^+_{k_z',k_z}}\bm z\times(\kk-\kk').
\end{align}
Inspection of Eq.~\eqref{eq:sjmomentresults} shows that due to the integral over $\kk'$ only the terms that contain $\bm z\times \kk$ in $\d \rr_{\kk\kk'}$ contribute to the magnetic moment. Using $W^{\textrm{S}}_{\kk\kk'}=\frac{2\pi}{\hbar}n_iV_0^2(\Gamma^+_{kz,kz'})^2\d(\e^+_{v\kk}-\e^+_{v\kk'})$, for the $z$-component of the side jump contribution we obtain
\begin{align}\label{eq:msj_appendix}
  m^{\textrm{sj}}_z&=2\pi e\tau n_i V_0^2\frac{ k_\perp^2}{m^v_\perp}\int (d\kk')\Gamma^+_{kz,kz'}\d(\e^+_{v\kk}-\e^+_{v\kk'})\nonumber\\
   &\left[\frac{\gamma_{k_z}\Gamma^+_{k_z',k_z}P_1^2}{[\e_{v}^+(\kk)-\e_{c}(\kk)]^2}+\frac{\gamma_{k_z}\Gamma^+_{k_z',k_z}P_4^2}{[\e_{v}^+(\kk)-\e_{h}(\kk)]^2}
+\frac{\Gamma^-_{k_z',k_z}P_1^2}{[\e_{v}^+(\kk')-\e_{c}(\kk')][\e_{v}^+(\kk)-\e_{c}(\kk)]}+
\frac{\Gamma^-_{k_z',k_z}P_4^2}{[\e_{v}^+(\kk')-\e_{h}(\kk')][\e_{v}^+(\kk)-\e_{h}(\kk)]}\right).
\end{align}
\twocolumngrid

\bibliography{extrinsic_references}

\begin{thebibliography}{46}
\expandafter\ifx\csname natexlab\endcsname\relax\def\natexlab#1{#1}\fi
\expandafter\ifx\csname bibnamefont\endcsname\relax
  \def\bibnamefont#1{#1}\fi
\expandafter\ifx\csname bibfnamefont\endcsname\relax
  \def\bibfnamefont#1{#1}\fi
\expandafter\ifx\csname citenamefont\endcsname\relax
  \def\citenamefont#1{#1}\fi
\expandafter\ifx\csname url\endcsname\relax
  \def\url#1{\texttt{#1}}\fi
\expandafter\ifx\csname urlprefix\endcsname\relax\def\urlprefix{URL }\fi
\providecommand{\bibinfo}[2]{#2}
\providecommand{\eprint}[2][]{\url{#2}}

\bibitem[{\citenamefont{Gambardella and Miron}(2011)}]{Gambardella2011}
\bibinfo{author}{\bibfnamefont{P.}~\bibnamefont{Gambardella}} \bibnamefont{and}
  \bibinfo{author}{\bibfnamefont{I.~M.} \bibnamefont{Miron}},
  \bibinfo{journal}{Philos. T. R. Soc. S-A} \textbf{\bibinfo{volume}{369}},
  \bibinfo{pages}{3175} (\bibinfo{year}{2011}).

\bibitem[{\citenamefont{Kato et~al.}(2004)\citenamefont{Kato, Myers, Gossard,
  and Awschalom}}]{Kato2004magnetization}
\bibinfo{author}{\bibfnamefont{Y.~K.} \bibnamefont{Kato}},
  \bibinfo{author}{\bibfnamefont{R.~C.} \bibnamefont{Myers}},
  \bibinfo{author}{\bibfnamefont{A.~C.} \bibnamefont{Gossard}},
  \bibnamefont{and} \bibinfo{author}{\bibfnamefont{D.~D.}
  \bibnamefont{Awschalom}}, \bibinfo{journal}{Phys. Rev. Lett.}
  \textbf{\bibinfo{volume}{93}}, \bibinfo{pages}{176601}
  (\bibinfo{year}{2004}).

\bibitem[{\citenamefont{Silov et~al.}(2004)\citenamefont{Silov, Blajnov,
  Wolter, Hey, Ploog, and Averkiev}}]{silov2004current}
\bibinfo{author}{\bibfnamefont{A.~Y.} \bibnamefont{Silov}},
  \bibinfo{author}{\bibfnamefont{P.}~\bibnamefont{Blajnov}},
  \bibinfo{author}{\bibfnamefont{J.}~\bibnamefont{Wolter}},
  \bibinfo{author}{\bibfnamefont{R.}~\bibnamefont{Hey}},
  \bibinfo{author}{\bibfnamefont{K.}~\bibnamefont{Ploog}}, \bibnamefont{and}
  \bibinfo{author}{\bibfnamefont{N.}~\bibnamefont{Averkiev}},
  \bibinfo{journal}{Appl. phys. lett.} \textbf{\bibinfo{volume}{85}},
  \bibinfo{pages}{5929} (\bibinfo{year}{2004}).

\bibitem[{\citenamefont{Kato et~al.}(2005)\citenamefont{Kato, Myers, Gossard,
  and Awschalom}}]{Kato2005magnetization}
\bibinfo{author}{\bibfnamefont{Y.}~\bibnamefont{Kato}},
  \bibinfo{author}{\bibfnamefont{R.}~\bibnamefont{Myers}},
  \bibinfo{author}{\bibfnamefont{A.}~\bibnamefont{Gossard}}, \bibnamefont{and}
  \bibinfo{author}{\bibfnamefont{D.}~\bibnamefont{Awschalom}},
  \bibinfo{journal}{Appl. Phys. Lett.} \textbf{\bibinfo{volume}{87}},
  \bibinfo{pages}{022503} (\bibinfo{year}{2005}).

\bibitem[{\citenamefont{Sih et~al.}(2005)\citenamefont{Sih, Myers, Kato, Lau,
  Gossard, and Awschalom}}]{Sih2005}
\bibinfo{author}{\bibfnamefont{V.}~\bibnamefont{Sih}},
  \bibinfo{author}{\bibfnamefont{R.}~\bibnamefont{Myers}},
  \bibinfo{author}{\bibfnamefont{Y.}~\bibnamefont{Kato}},
  \bibinfo{author}{\bibfnamefont{W.}~\bibnamefont{Lau}},
  \bibinfo{author}{\bibfnamefont{A.}~\bibnamefont{Gossard}}, \bibnamefont{and}
  \bibinfo{author}{\bibfnamefont{D.}~\bibnamefont{Awschalom}},
  \bibinfo{journal}{Nat. Phys.} \textbf{\bibinfo{volume}{1}},
  \bibinfo{pages}{31} (\bibinfo{year}{2005}).

\bibitem[{\citenamefont{Stern et~al.}(2006)\citenamefont{Stern, Ghosh, Xiang,
  Zhu, Samarth, and Awschalom}}]{Stern2006}
\bibinfo{author}{\bibfnamefont{N.~P.} \bibnamefont{Stern}},
  \bibinfo{author}{\bibfnamefont{S.}~\bibnamefont{Ghosh}},
  \bibinfo{author}{\bibfnamefont{G.}~\bibnamefont{Xiang}},
  \bibinfo{author}{\bibfnamefont{M.}~\bibnamefont{Zhu}},
  \bibinfo{author}{\bibfnamefont{N.}~\bibnamefont{Samarth}}, \bibnamefont{and}
  \bibinfo{author}{\bibfnamefont{D.~D.} \bibnamefont{Awschalom}},
  \bibinfo{journal}{Phys. Rev. Lett.} \textbf{\bibinfo{volume}{97}},
  \bibinfo{pages}{126603} (\bibinfo{year}{2006}).

\bibitem[{\citenamefont{Yang et~al.}(2006)\citenamefont{Yang, He, Ding, Cui,
  Zeng, Wang, and Ge}}]{Yang2006}
\bibinfo{author}{\bibfnamefont{C.}~\bibnamefont{Yang}},
  \bibinfo{author}{\bibfnamefont{H.}~\bibnamefont{He}},
  \bibinfo{author}{\bibfnamefont{L.}~\bibnamefont{Ding}},
  \bibinfo{author}{\bibfnamefont{L.}~\bibnamefont{Cui}},
  \bibinfo{author}{\bibfnamefont{Y.}~\bibnamefont{Zeng}},
  \bibinfo{author}{\bibfnamefont{J.}~\bibnamefont{Wang}}, \bibnamefont{and}
  \bibinfo{author}{\bibfnamefont{W.}~\bibnamefont{Ge}}, \bibinfo{journal}{Phys.
  Rev. Lett.} \textbf{\bibinfo{volume}{96}}, \bibinfo{pages}{186605}
  (\bibinfo{year}{2006}).

\bibitem[{\citenamefont{Levitov et~al.}(1985)\citenamefont{Levitov, Nazarov,
  and Eliashberg}}]{Levitov}
\bibinfo{author}{\bibfnamefont{L.~S.} \bibnamefont{Levitov}},
  \bibinfo{author}{\bibfnamefont{Y.~V.} \bibnamefont{Nazarov}},
  \bibnamefont{and} \bibinfo{author}{\bibfnamefont{G.~M.}
  \bibnamefont{Eliashberg}}, \bibinfo{journal}{Sov. Phys. JETP}
  \textbf{\bibinfo{volume}{61}}, \bibinfo{pages}{133} (\bibinfo{year}{1985}).

\bibitem[{\citenamefont{Yoda et~al.}(2015)\citenamefont{Yoda, Yokoyama, and
  Murakami}}]{Yoda2015}
\bibinfo{author}{\bibfnamefont{T.}~\bibnamefont{Yoda}},
  \bibinfo{author}{\bibfnamefont{T.}~\bibnamefont{Yokoyama}}, \bibnamefont{and}
  \bibinfo{author}{\bibfnamefont{S.}~\bibnamefont{Murakami}},
  \bibinfo{journal}{Sci.rep.} \textbf{\bibinfo{volume}{5}},
  \bibinfo{pages}{12024} (\bibinfo{year}{2015}).

\bibitem[{\citenamefont{Zhong et~al.}(2016)\citenamefont{Zhong, Moore, and
  Souza}}]{Zhong2016}
\bibinfo{author}{\bibfnamefont{S.}~\bibnamefont{Zhong}},
  \bibinfo{author}{\bibfnamefont{J.~E.} \bibnamefont{Moore}}, \bibnamefont{and}
  \bibinfo{author}{\bibfnamefont{I.}~\bibnamefont{Souza}},
  \bibinfo{journal}{Phys. Rev. Lett.} \textbf{\bibinfo{volume}{116}},
  \bibinfo{pages}{077201} (\bibinfo{year}{2016}).

\bibitem[{\citenamefont{Shalygin et~al.}(2012)\citenamefont{Shalygin, Sofronov,
  Vorob'ev, and Farbshtein}}]{Farbshtein2012}
\bibinfo{author}{\bibfnamefont{V.}~\bibnamefont{Shalygin}},
  \bibinfo{author}{\bibfnamefont{A.}~\bibnamefont{Sofronov}},
  \bibinfo{author}{\bibfnamefont{L.}~\bibnamefont{Vorob'ev}}, \bibnamefont{and}
  \bibinfo{author}{\bibfnamefont{I.}~\bibnamefont{Farbshtein}},
  \bibinfo{journal}{Phys. Solid State} \textbf{\bibinfo{volume}{54}},
  \bibinfo{pages}{2362} (\bibinfo{year}{2012}).

\bibitem[{\citenamefont{Vorob'ev et~al.}(1979)\citenamefont{Vorob'ev, Ivchenko,
  Pikus, Farbshtein, Shalygin, and Shturbin}}]{Ivchenko1979}
\bibinfo{author}{\bibfnamefont{L.}~\bibnamefont{Vorob'ev}},
  \bibinfo{author}{\bibfnamefont{E.}~\bibnamefont{Ivchenko}},
  \bibinfo{author}{\bibfnamefont{G.}~\bibnamefont{Pikus}},
  \bibinfo{author}{\bibfnamefont{I.}~\bibnamefont{Farbshtein}},
  \bibinfo{author}{\bibfnamefont{V.}~\bibnamefont{Shalygin}}, \bibnamefont{and}
  \bibinfo{author}{\bibfnamefont{A.}~\bibnamefont{Shturbin}},
  \bibinfo{journal}{Sov. Phys. JETP} \textbf{\bibinfo{volume}{29}},
  \bibinfo{pages}{441} (\bibinfo{year}{1979}).

\bibitem[{\citenamefont{Ivchenko and Pikus}(1978)}]{Ivchenko1978}
\bibinfo{author}{\bibfnamefont{E.}~\bibnamefont{Ivchenko}} \bibnamefont{and}
  \bibinfo{author}{\bibfnamefont{G.}~\bibnamefont{Pikus}},
  \bibinfo{journal}{Sov. Phys. JETP Lett.} \textbf{\bibinfo{volume}{27}},
  \bibinfo{pages}{604} (\bibinfo{year}{1978}).

\bibitem[{\citenamefont{Vasko and Prima}(1979)}]{VaskoPrima1979}
\bibinfo{author}{\bibfnamefont{F.}~\bibnamefont{Vasko}} \bibnamefont{and}
  \bibinfo{author}{\bibfnamefont{N.}~\bibnamefont{Prima}},
  \bibinfo{journal}{Sov. Phys. Solid State} \textbf{\bibinfo{volume}{21}},
  \bibinfo{pages}{994} (\bibinfo{year}{1979}).

\bibitem[{\citenamefont{Edelstein}(1990)}]{Edelstein1990}
\bibinfo{author}{\bibfnamefont{V.~M.} \bibnamefont{Edelstein}},
  \bibinfo{journal}{Solid State Commun.} \textbf{\bibinfo{volume}{73}},
  \bibinfo{pages}{233} (\bibinfo{year}{1990}).

\bibitem[{\citenamefont{Aronov et~al.}(1991)\citenamefont{Aronov,
  Lyanda-Geller, Pikus, and Parsons}}]{Aronov1991}
\bibinfo{author}{\bibfnamefont{A.}~\bibnamefont{Aronov}},
  \bibinfo{author}{\bibfnamefont{Y.~B.} \bibnamefont{Lyanda-Geller}},
  \bibinfo{author}{\bibfnamefont{G.}~\bibnamefont{Pikus}}, \bibnamefont{and}
  \bibinfo{author}{\bibfnamefont{D.}~\bibnamefont{Parsons}},
  \bibinfo{journal}{Sov. Phys. JETP} \textbf{\bibinfo{volume}{73}},
  \bibinfo{pages}{537} (\bibinfo{year}{1991}).

\bibitem[{\citenamefont{Ganichev et~al.}(2016)\citenamefont{Ganichev, Trushin,
  and Schliemann}}]{ganichev2016}
\bibinfo{author}{\bibfnamefont{S.~D.} \bibnamefont{Ganichev}},
  \bibinfo{author}{\bibfnamefont{M.}~\bibnamefont{Trushin}}, \bibnamefont{and}
  \bibinfo{author}{\bibfnamefont{J.}~\bibnamefont{Schliemann}},
  \bibinfo{journal}{arXiv:1606.02043}  (\bibinfo{year}{2016}).

\bibitem[{\citenamefont{Rou et~al.}(2017)\citenamefont{Rou,
  \ifmmode~\mbox{\c{S}}\else \c{S}\fi{}ahin, Ma, and Pesin}}]{Rou2017}
\bibinfo{author}{\bibfnamefont{J.}~\bibnamefont{Rou}},
  \bibinfo{author}{\bibfnamefont{C.}~\bibnamefont{\ifmmode~\mbox{\c{S}}\else
  \c{S}\fi{}ahin}}, \bibinfo{author}{\bibfnamefont{J.}~\bibnamefont{Ma}},
  \bibnamefont{and} \bibinfo{author}{\bibfnamefont{D.~A.} \bibnamefont{Pesin}},
  \bibinfo{journal}{Phys. Rev. B} \textbf{\bibinfo{volume}{96}},
  \bibinfo{pages}{035120} (\bibinfo{year}{2017}).

\bibitem[{\citenamefont{Nagaosa et~al.}(2010)\citenamefont{Nagaosa, Sinova,
  Onoda, MacDonald, and Ong}}]{NagaosaReview}
\bibinfo{author}{\bibfnamefont{N.}~\bibnamefont{Nagaosa}},
  \bibinfo{author}{\bibfnamefont{J.}~\bibnamefont{Sinova}},
  \bibinfo{author}{\bibfnamefont{S.}~\bibnamefont{Onoda}},
  \bibinfo{author}{\bibfnamefont{A.~H.} \bibnamefont{MacDonald}},
  \bibnamefont{and} \bibinfo{author}{\bibfnamefont{N.~P.} \bibnamefont{Ong}},
  \bibinfo{journal}{Rev. Mod. Phys.} \textbf{\bibinfo{volume}{82}},
  \bibinfo{pages}{1539} (\bibinfo{year}{2010}).

\bibitem[{foo()}]{footnoteHE}
\bibinfo{note}{We would like to comment on the absence of a Berry curvature
  contribution to the KME, see Eq.~\eqref{eq:KME} of the main text, in finite
  samples. In equilibrium, there is a well-known contribution to the
  magnetization produced by the Berry curvature, which can be shown to be
  produced by boundary Hall currents in a crystal with broken time-reversal
  symmetry (see Ref.~\onlinecite{XiaoNiu} for a review of relevant literature).
  In the present case, such equilibrium magnetization vanishes due to time
  reversal invariance. However, if (bulk) electrons are driven out of
  equilibrium near a crystal's boundary by a transport electric field,
  anomalous velocities of electrons in the confining and transport electric
  fields do produce an additional local conductivity tensor. For a band with
  energy $\e_n$, Berry curvature $\bm \Omega_{n}$, and space-dependent
  equilibrium occupation $f_n(\rr)$, such boundary conductivity tensor is given
  by\cite{Sahinunpublished} \begin{align} \sigma_{ab}^{\textrm{bndr}}=-\tau
  e^2\int
  (d\kk)(\e_{acd}\p_b\e_n+\e_{bcd}\p_a\e_n)\Omega_{n,d}\p_{\rr_c}f_n(\rr).
  \end{align} The tensor is purely symmetric. At first, the current due to this
  boundary conductivity tensor appears to be able to produce bulk
  magnetization. However, the existence of such magnetization would lead to a
  logical contradiction. The contradiction stems from the fact that the KME is
  the Onsager-reciprocal effect of the natural optical
  activity\cite{Levitov,Zhong2016,Rou2017}. Indeed, let us imagine there is a
  bulk magnetization induced by a transport electric field due to the boundary
  conductivity tensor above. If the transport electric field is slightly
  non-uniform in the bulk of the crystal, such magnetization will give rise to
  a ``curl of magnetization" bulk current. That current \textit{must} be a part
  of the natural optical activity tensor in a time-reversal invariant metal due
  to Onsager relations\cite{MaPesin2015}. That is, one is forced to conclude
  there is an additional Berry-curvature-related contribution to the natural
  optical activity tensor. However, it is well known\cite{AgranovichYudson}
  that natural optical activity necessarily implies the existence of an
  \textit{antisymmetric} conductivity tensor at the boundary, which ensures the
  absence of the polar Kerr effect in a time reversal invariant
  sample\cite{HosurKerr2014}. Finally, since the Berry-curvature related
  boundary conductivity tensor does \textit{not} have an antisymmetric part,
  the above argument implies that the corresponding contribution to the natural
  optical activity simply does not exist, and Eq.~\eqref{eq:KME} stands as
  describing the total KME. We leave it for future work to determine the exact
  mechanism of cancelation of magnetization due to the Berry-curvature related
  boundary current; such cancelation can possibly be related the surface states
  of the crystal.}

\bibitem[{\citenamefont{Xiao et~al.}(2010)\citenamefont{Xiao, Chang, and
  Niu}}]{XiaoNiu}
\bibinfo{author}{\bibfnamefont{D.}~\bibnamefont{Xiao}},
  \bibinfo{author}{\bibfnamefont{M.-C.} \bibnamefont{Chang}}, \bibnamefont{and}
  \bibinfo{author}{\bibfnamefont{Q.}~\bibnamefont{Niu}}, \bibinfo{journal}{Rev.
  Mod. Phys.} \textbf{\bibinfo{volume}{82}}, \bibinfo{pages}{1959}
  (\bibinfo{year}{2010}).

\bibitem[{\citenamefont{Luttinger}(1958)}]{Luttinger1958}
\bibinfo{author}{\bibfnamefont{J.~M.} \bibnamefont{Luttinger}},
  \bibinfo{journal}{Phys. Rev.} \textbf{\bibinfo{volume}{112}},
  \bibinfo{pages}{739} (\bibinfo{year}{1958}).

\bibitem[{\citenamefont{Leroux-Hugon and Ghazali}(1972)}]{Ghazali1972}
\bibinfo{author}{\bibfnamefont{P.}~\bibnamefont{Leroux-Hugon}}
  \bibnamefont{and} \bibinfo{author}{\bibfnamefont{A.}~\bibnamefont{Ghazali}},
  \bibinfo{journal}{J. Phys. C} \textbf{\bibinfo{volume}{5}},
  \bibinfo{pages}{1072} (\bibinfo{year}{1972}).

\bibitem[{\citenamefont{Belinicher et~al.}(1982)\citenamefont{Belinicher,
  Ivchenko, and Sturman}}]{Belinicher1982}
\bibinfo{author}{\bibfnamefont{V.~I.} \bibnamefont{Belinicher}},
  \bibinfo{author}{\bibfnamefont{E.~L.} \bibnamefont{Ivchenko}},
  \bibnamefont{and} \bibinfo{author}{\bibfnamefont{B.~I.}
  \bibnamefont{Sturman}}, \bibinfo{journal}{Sov. Phys. JETP}
  \textbf{\bibinfo{volume}{56}}, \bibinfo{pages}{359} (\bibinfo{year}{1982}).

\bibitem[{\citenamefont{Sinitsyn et~al.}(2006)\citenamefont{Sinitsyn, Niu, and
  MacDonald}}]{Sinitsyn2006}
\bibinfo{author}{\bibfnamefont{N.~A.} \bibnamefont{Sinitsyn}},
  \bibinfo{author}{\bibfnamefont{Q.}~\bibnamefont{Niu}}, \bibnamefont{and}
  \bibinfo{author}{\bibfnamefont{A.~H.} \bibnamefont{MacDonald}},
  \bibinfo{journal}{Phys. Rev. B} \textbf{\bibinfo{volume}{73}},
  \bibinfo{pages}{075318} (\bibinfo{year}{2006}).

\bibitem[{\citenamefont{Nomura}(1960)}]{Nomura1960}
\bibinfo{author}{\bibfnamefont{K.~C.} \bibnamefont{Nomura}},
  \bibinfo{journal}{Phys. Rev. Lett.} \textbf{\bibinfo{volume}{5}},
  \bibinfo{pages}{500} (\bibinfo{year}{1960}).

\bibitem[{\citenamefont{Ivchenko and Pikus}(1975)}]{IvchenkoTe}
\bibinfo{author}{\bibfnamefont{E.}~\bibnamefont{Ivchenko}} \bibnamefont{and}
  \bibinfo{author}{\bibfnamefont{G.}~\bibnamefont{Pikus}},
  \bibinfo{journal}{Sov. Phys. Solid State} \textbf{\bibinfo{volume}{16}},
  \bibinfo{pages}{1261} (\bibinfo{year}{1975}).

\bibitem[{\citenamefont{Agapito et~al.}(2013)\citenamefont{Agapito, Kioussis,
  Goddard~III, and Ong}}]{Agapito2013}
\bibinfo{author}{\bibfnamefont{L.~A.} \bibnamefont{Agapito}},
  \bibinfo{author}{\bibfnamefont{N.}~\bibnamefont{Kioussis}},
  \bibinfo{author}{\bibfnamefont{W.~A.} \bibnamefont{Goddard~III}},
  \bibnamefont{and} \bibinfo{author}{\bibfnamefont{N.}~\bibnamefont{Ong}},
  \bibinfo{journal}{Phys. Rev. Lett.} \textbf{\bibinfo{volume}{110}},
  \bibinfo{pages}{176401} (\bibinfo{year}{2013}).

\bibitem[{\citenamefont{Hirayama et~al.}(2015)\citenamefont{Hirayama, Okugawa,
  Ishibashi, Murakami, and Miyake}}]{Hirayama2015}
\bibinfo{author}{\bibfnamefont{M.}~\bibnamefont{Hirayama}},
  \bibinfo{author}{\bibfnamefont{R.}~\bibnamefont{Okugawa}},
  \bibinfo{author}{\bibfnamefont{S.}~\bibnamefont{Ishibashi}},
  \bibinfo{author}{\bibfnamefont{S.}~\bibnamefont{Murakami}}, \bibnamefont{and}
  \bibinfo{author}{\bibfnamefont{T.}~\bibnamefont{Miyake}},
  \bibinfo{journal}{Phys. Rev. Lett.} \textbf{\bibinfo{volume}{114}},
  \bibinfo{pages}{206401} (\bibinfo{year}{2015}).

\bibitem[{\citenamefont{Nakayama et~al.}(2017)\citenamefont{Nakayama, Kuno,
  Yamauchi, Souma, Sugawara, Oguchi, Sato, and Takahashi}}]{Nakayama2017}
\bibinfo{author}{\bibfnamefont{K.}~\bibnamefont{Nakayama}},
  \bibinfo{author}{\bibfnamefont{M.}~\bibnamefont{Kuno}},
  \bibinfo{author}{\bibfnamefont{K.}~\bibnamefont{Yamauchi}},
  \bibinfo{author}{\bibfnamefont{S.}~\bibnamefont{Souma}},
  \bibinfo{author}{\bibfnamefont{K.}~\bibnamefont{Sugawara}},
  \bibinfo{author}{\bibfnamefont{T.}~\bibnamefont{Oguchi}},
  \bibinfo{author}{\bibfnamefont{T.}~\bibnamefont{Sato}}, \bibnamefont{and}
  \bibinfo{author}{\bibfnamefont{T.}~\bibnamefont{Takahashi}},
  \bibinfo{journal}{Phys. Rev. B} \textbf{\bibinfo{volume}{95}},
  \bibinfo{pages}{125204} (\bibinfo{year}{2017}).

\bibitem[{\citenamefont{Doi et~al.}(1970)\citenamefont{Doi, Nakao, and
  Kamimura}}]{Doi1970}
\bibinfo{author}{\bibfnamefont{T.}~\bibnamefont{Doi}},
  \bibinfo{author}{\bibfnamefont{K.}~\bibnamefont{Nakao}}, \bibnamefont{and}
  \bibinfo{author}{\bibfnamefont{H.}~\bibnamefont{Kamimura}},
  \bibinfo{journal}{J. Phys. Soc. Jpn.} \textbf{\bibinfo{volume}{28}},
  \bibinfo{pages}{36} (\bibinfo{year}{1970}).

\bibitem[{\citenamefont{Joannopoulos et~al.}(1975)\citenamefont{Joannopoulos,
  Schl{\"u}ter, and Cohen}}]{Joannopoulos1975}
\bibinfo{author}{\bibfnamefont{J.}~\bibnamefont{Joannopoulos}},
  \bibinfo{author}{\bibfnamefont{M.}~\bibnamefont{Schl{\"u}ter}},
  \bibnamefont{and} \bibinfo{author}{\bibfnamefont{M.~L.} \bibnamefont{Cohen}},
  \bibinfo{journal}{Phys. Rev. B} \textbf{\bibinfo{volume}{11}},
  \bibinfo{pages}{2186} (\bibinfo{year}{1975}).

\bibitem[{\citenamefont{Reitz}(1957)}]{Reitz1957}
\bibinfo{author}{\bibfnamefont{J.~R.} \bibnamefont{Reitz}},
  \bibinfo{journal}{Physical Review} \textbf{\bibinfo{volume}{105}},
  \bibinfo{pages}{1233} (\bibinfo{year}{1957}).

\bibitem[{\citenamefont{Hartmann and Mahanti}(1972)}]{HartMann1972}
\bibinfo{author}{\bibfnamefont{W.}~\bibnamefont{Hartmann}} \bibnamefont{and}
  \bibinfo{author}{\bibfnamefont{S.}~\bibnamefont{Mahanti}},
  \bibinfo{journal}{J. Non. Cryst. Solids} \textbf{\bibinfo{volume}{8}},
  \bibinfo{pages}{633} (\bibinfo{year}{1972}).

\bibitem[{\citenamefont{Li et~al.}(2013)\citenamefont{Li, Ciani, Gayles,
  Papaconstantopoulos, Kioussis, Grein, and Aqariden}}]{Li2013}
\bibinfo{author}{\bibfnamefont{J.}~\bibnamefont{Li}},
  \bibinfo{author}{\bibfnamefont{A.}~\bibnamefont{Ciani}},
  \bibinfo{author}{\bibfnamefont{J.}~\bibnamefont{Gayles}},
  \bibinfo{author}{\bibfnamefont{D.}~\bibnamefont{Papaconstantopoulos}},
  \bibinfo{author}{\bibfnamefont{N.}~\bibnamefont{Kioussis}},
  \bibinfo{author}{\bibfnamefont{C.}~\bibnamefont{Grein}}, \bibnamefont{and}
  \bibinfo{author}{\bibfnamefont{F.}~\bibnamefont{Aqariden}},
  \bibinfo{journal}{Philos. Mag.} \textbf{\bibinfo{volume}{93}},
  \bibinfo{pages}{3216} (\bibinfo{year}{2013}).

\bibitem[{\citenamefont{Molina and Lomba}(2003)}]{Molina2003}
\bibinfo{author}{\bibfnamefont{D.}~\bibnamefont{Molina}} \bibnamefont{and}
  \bibinfo{author}{\bibfnamefont{E.}~\bibnamefont{Lomba}},
  \bibinfo{journal}{Phys. Rev. B} \textbf{\bibinfo{volume}{67}},
  \bibinfo{pages}{094208} (\bibinfo{year}{2003}).

\bibitem[{\citenamefont{Bresler et~al.}(1970)\citenamefont{Bresler, Veselago,
  Kosichkin, Pikus, Farbshtein, and Shalyt}}]{Bresler1970}
\bibinfo{author}{\bibfnamefont{M.}~\bibnamefont{Bresler}},
  \bibinfo{author}{\bibfnamefont{V.}~\bibnamefont{Veselago}},
  \bibinfo{author}{\bibfnamefont{Y.~V.} \bibnamefont{Kosichkin}},
  \bibinfo{author}{\bibfnamefont{G.}~\bibnamefont{Pikus}},
  \bibinfo{author}{\bibfnamefont{I.}~\bibnamefont{Farbshtein}},
  \bibnamefont{and} \bibinfo{author}{\bibfnamefont{S.}~\bibnamefont{Shalyt}},
  \bibinfo{journal}{Sov. Phys. JETP} \textbf{\bibinfo{volume}{30}},
  \bibinfo{pages}{799} (\bibinfo{year}{1970}).

\bibitem[{\citenamefont{Shinno et~al.}(1973)\citenamefont{Shinno, Yoshizaki,
  Tanaka, Doi, and Kamimura}}]{Shinno1973}
\bibinfo{author}{\bibfnamefont{H.}~\bibnamefont{Shinno}},
  \bibinfo{author}{\bibfnamefont{R.}~\bibnamefont{Yoshizaki}},
  \bibinfo{author}{\bibfnamefont{S.}~\bibnamefont{Tanaka}},
  \bibinfo{author}{\bibfnamefont{T.}~\bibnamefont{Doi}}, \bibnamefont{and}
  \bibinfo{author}{\bibfnamefont{H.}~\bibnamefont{Kamimura}},
  \bibinfo{journal}{J. Phys. Soc. Jpn.} \textbf{\bibinfo{volume}{35}},
  \bibinfo{pages}{525} (\bibinfo{year}{1973}).

\bibitem[{\citenamefont{Ma and Pesin}(2015)}]{MaPesin2015}
\bibinfo{author}{\bibfnamefont{J.}~\bibnamefont{Ma}} \bibnamefont{and}
  \bibinfo{author}{\bibfnamefont{D.~A.} \bibnamefont{Pesin}},
  \bibinfo{journal}{Phys. Rev. B} \textbf{\bibinfo{volume}{92}},
  \bibinfo{pages}{235205} (\bibinfo{year}{2015}).

\bibitem[{\citenamefont{Regel et~al.}(2001)\citenamefont{Regel, Parfeniev,
  Farbshtein, Shulpina, Yakimov, Shalimov, and Turchaninov}}]{regel2001defect}
\bibinfo{author}{\bibfnamefont{L.}~\bibnamefont{Regel}},
  \bibinfo{author}{\bibfnamefont{R.}~\bibnamefont{Parfeniev}},
  \bibinfo{author}{\bibfnamefont{I.}~\bibnamefont{Farbshtein}},
  \bibinfo{author}{\bibfnamefont{I.}~\bibnamefont{Shulpina}},
  \bibinfo{author}{\bibfnamefont{S.}~\bibnamefont{Yakimov}},
  \bibinfo{author}{\bibfnamefont{V.}~\bibnamefont{Shalimov}}, \bibnamefont{and}
  \bibinfo{author}{\bibfnamefont{A.}~\bibnamefont{Turchaninov}},
  \bibinfo{journal}{Processing by Centrifugation} p. \bibinfo{pages}{241}
  (\bibinfo{year}{2001}).

\bibitem[{\citenamefont{Bresler et~al.}(1972)\citenamefont{Bresler, Dolgopolov,
  Poltoratskij, Svechkarev, and Farbshtein}}]{Bresler1972}
\bibinfo{author}{\bibfnamefont{M.}~\bibnamefont{Bresler}},
  \bibinfo{author}{\bibfnamefont{D.}~\bibnamefont{Dolgopolov}},
  \bibinfo{author}{\bibfnamefont{V.}~\bibnamefont{Poltoratskij}},
  \bibinfo{author}{\bibfnamefont{I.}~\bibnamefont{Svechkarev}},
  \bibnamefont{and}
  \bibinfo{author}{\bibfnamefont{I.}~\bibnamefont{Farbshtein}},
  \bibinfo{journal}{Sov. Phys. JETP} \textbf{\bibinfo{volume}{35}},
  \bibinfo{pages}{150} (\bibinfo{year}{1972}).

\bibitem[{\citenamefont{Chernyshov et~al.}(2009)\citenamefont{Chernyshov,
  Overby, Liu, Furdyna, Lyanda-Geller, and Rokhinson}}]{Chernyshov2009}
\bibinfo{author}{\bibfnamefont{A.}~\bibnamefont{Chernyshov}},
  \bibinfo{author}{\bibfnamefont{M.}~\bibnamefont{Overby}},
  \bibinfo{author}{\bibfnamefont{X.}~\bibnamefont{Liu}},
  \bibinfo{author}{\bibfnamefont{J.~K.} \bibnamefont{Furdyna}},
  \bibinfo{author}{\bibfnamefont{Y.}~\bibnamefont{Lyanda-Geller}},
  \bibnamefont{and} \bibinfo{author}{\bibfnamefont{L.~P.}
  \bibnamefont{Rokhinson}}, \bibinfo{journal}{Nature Phys.}
  \textbf{\bibinfo{volume}{5}}, \bibinfo{pages}{656} (\bibinfo{year}{2009}).

\bibitem[{\citenamefont{Tsirkin et~al.}(2017)\citenamefont{Tsirkin, Puente, and
  Souza}}]{tsirkin2017}
\bibinfo{author}{\bibfnamefont{S.~S.} \bibnamefont{Tsirkin}},
  \bibinfo{author}{\bibfnamefont{P.~A.} \bibnamefont{Puente}},
  \bibnamefont{and} \bibinfo{author}{\bibfnamefont{I.}~\bibnamefont{Souza}},
  \bibinfo{journal}{arXiv:1710.03204}  (\bibinfo{year}{2017}).

\bibitem[{\citenamefont{\c{S}ahin et~al.}()\citenamefont{\c{S}ahin, Rou, Ma,
  and Pesin}}]{Sahinunpublished}
\bibinfo{author}{\bibfnamefont{C.}~\bibnamefont{\c{S}ahin}},
  \bibinfo{author}{\bibfnamefont{J.}~\bibnamefont{Rou}},
  \bibinfo{author}{\bibfnamefont{J.}~\bibnamefont{Ma}}, \bibnamefont{and}
  \bibinfo{author}{\bibfnamefont{D.~A.} \bibnamefont{Pesin}},
  \bibinfo{note}{unpublished}.

\bibitem[{\citenamefont{Agranovich and Yudson}(1973)}]{AgranovichYudson}
\bibinfo{author}{\bibfnamefont{V.}~\bibnamefont{Agranovich}} \bibnamefont{and}
  \bibinfo{author}{\bibfnamefont{V.}~\bibnamefont{Yudson}},
  \bibinfo{journal}{Opt. Commun.} \textbf{\bibinfo{volume}{9}},
  \bibinfo{pages}{58 } (\bibinfo{year}{1973}).

\bibitem[{\citenamefont{Hosur et~al.}(2015)\citenamefont{Hosur, Kapitulnik,
  Kivelson, Orenstein, Raghu, Cho, and Fried}}]{HosurKerr2014}
\bibinfo{author}{\bibfnamefont{P.}~\bibnamefont{Hosur}},
  \bibinfo{author}{\bibfnamefont{A.}~\bibnamefont{Kapitulnik}},
  \bibinfo{author}{\bibfnamefont{S.~A.} \bibnamefont{Kivelson}},
  \bibinfo{author}{\bibfnamefont{J.}~\bibnamefont{Orenstein}},
  \bibinfo{author}{\bibfnamefont{S.}~\bibnamefont{Raghu}},
  \bibinfo{author}{\bibfnamefont{W.}~\bibnamefont{Cho}}, \bibnamefont{and}
  \bibinfo{author}{\bibfnamefont{A.}~\bibnamefont{Fried}},
  \bibinfo{journal}{Phys. Rev. B} \textbf{\bibinfo{volume}{91}},
  \bibinfo{pages}{E039908} (\bibinfo{year}{2015}).

\end{thebibliography}
\bibliographystyle{apsrev}

\end{document}